\documentclass[useAMS,usenatbib]{mnras}
\usepackage{graphicx}
\usepackage{amsmath, amssymb}
\usepackage[T1]{fontenc}
\usepackage[usenames]{color}
\usepackage[dvipsnames]{xcolor}
\usepackage{bm}
\usepackage[capitalize]{cleveref}
\usepackage{physics}
\usepackage{acro}
\usepackage{caption}
\usepackage{subcaption}
\usepackage{algorithm, algpseudocode}

\DeclareRobustCommand{\VAN}[3]{#2}
\let\VANthebibliography\thebibliography
\def\thebibliography{\DeclareRobustCommand{\VAN}[3]{##3}\VANthebibliography}

\newcommand{\bibnote}[2]{\global\@namedef{#1note}{#2}}
\newcommand{\biblink}[2]{\global\@namedef{#1link}{#2}}

\def\LCDM/{$\Lambda$CDM}
\def\Vobs/{$V_{\text{obs}}$}
\def\Vbar/{$V_{\text{bar}}$}
\def\Vmond/{$V_{\text{MOND}}$}
\def\Vcdm/{$V_{\Lambda\text{CDM}}$}

\newcommand\HI{$\textrm{H}\scriptstyle\mathrm{I}$}

\title[Reassessing Renzo's Rule]{Renzo's rule revisited: A statistical study of galaxies' baryon--dark matter coupling}
\author[Ko et al.]{Enoch Ko\thanks{Email: \href{mailto:enoch.eko@gmail.com}{enoch.eko@gmail.com}}$^{1,2,3}$, Tariq Yasin$^1$, Harry Desmond$^4$, Richard Stiskalek$^1$ and Matt J. Jarvis$^{1,5}$
\\
$^{1}$Astrophysics, University of Oxford, Denys Wilkinson Building, Keble Road, Oxford, OX1 3RH, UK\\
$^{2}$DAMTP, Centre for Mathematical Sciences, University of Cambridge, Cambridge, CB3 0WA, UK\\
$^{3}$Department of Physics, University of Warwick, Coventry, CV4 7AL, UK\\
$^{4}$Institute of Cosmology \& Gravitation, University of Portsmouth, Dennis Sciama Building, Portsmouth, PO1 3FX, UK\\
$^{5}$Department of Physics and Astronomy, University of the Western Cape, Robert Sobukwe Road, 7535 Bellville, Cape Town, South Africa
}

\date{Accepted 2025 November 7. Received 2025 October 27; in original form 2025 August 6}

\begin{document}\label{firstpage}
\pagerange{\pageref{firstpage}--\pageref{lastpage}}
\maketitle

\begin{abstract}
We present a systematic statistical analysis of an informal astrophysical phenomenon known as \emph{Renzo's rule} (or \emph{Sancisi's law}), which states that ``for any feature in a galaxy’s luminosity profile, there is a corresponding feature in the rotation curve, and vice versa.''
This is often posed as a challenge for the standard \LCDM/ model while supporting alternative theories such as MOND.
Indeed, we identify clear features in the dwarf spiral NGC 1560 -- a prime example for Renzo's rule -- and find correlation statistics which support Renzo's rule with a slight preference for MOND over \LCDM/ halo fits.
However, a broader analysis on galaxies in the SPARC database reveals an excess of features in rotation curves that lack clear baryonic counterparts, with correlation statistics deviating up to $3\sigma$ on average from that predicted by both MOND and \LCDM/ haloes, challenging the validity of Renzo's rule.
Thus we do not find clear evidence for Renzo's rule in present galaxy data overall.
We additionally perform mock tests, which show that a definitive test of Renzo's rule is primarily limited by the lack of clearly resolved baryonic features in current galaxy data.
\end{abstract}


\section{Introduction}\label{sec:intro}
In the 1980s, observations of galaxy rotation curves (RCs) by \cite{1980ApJ...238..471R} revealed a significant discrepancy between the observed dynamics of galaxies and predictions from General Relativity based on their visible mass-energy content. This paved the way for the wide acceptance of today's standard model of cosmology --- the $\Lambda$ Cold Dark Matter (\LCDM/) model.
In \LCDM/, the majority of our Universe’s mass comes in the form of a non-baryonic component known as dark matter (DM), which only interacts gravitationally with both itself and ordinary matter.

Nevertheless, despite decades of extensive research and remarkable observational advances, DM has not been detected and many facets of galaxy kinematics remain mysterious (for reviews, see~\citealt{DelPopolo_review,Bullock_review,McGaugh_tale}).
Here, we investigate one such facet known as \emph{Renzo’s rule} (\citealt{Sancisi:2003xt}; also known as \emph{Sancisi's law}), which states that

\vspace{1mm}
\emph{``For any feature in a galaxy’s luminosity profile, there is a corresponding feature in the RC, and vice versa.''}
\vspace{1mm}

\noindent In \LCDM/, this is expected in regions where luminous stars dominate the dynamics, which would naturally reflect features in their distribution in the total gravitational potential.
However, the rule does not restrict itself to high-surface-brightness regions, but rather purports to describe a regularity \emph{anywhere} along a rotation curve.
In particular, some studies, e.g., \cite{Famaey_2012} and \cite{McGaugh_2019}, have stressed that Renzo's rule seems to apply even in the presence of a large discrepancy between the visible and total dynamical mass, i.e., regions dominated by DM according to the \LCDM/ model.
This would argue for a strong correlation between luminous and dark matter, which is peculiar given the assumed collisionless nature of the latter: the DM halo is dynamically hot and pressure supported, so should not exhibit the same features as galaxies.

A more natural explanation for the putative rule is given by Modified Newtonian Dynamics (MOND; \citealt{1983ApJ...270..365M, 1983ApJ...270..371M, 1983ApJ...270..384M}).
According to MOND, the kinematic discrepancies in RCs are not due to the existence of DM haloes, but are rather the result of a modified Newtonian gravitational field,
\begin{equation}\label{eqn:MOND}
    \bm{g} = \bm{g_N}\;\nu\left(\frac{g_N}{a_0}\right).
\end{equation}
Here, $\bm{g}$ is the total gravitational field, $\bm{g_N}$ is the Newtonian gravitational field due to baryons with strength $g_N = \vert\bm{g_N}\vert$, $a_0 \approx 1.2\times10^{-10}$ ms$^{-2}$ is a new fundamental constant --- a critical acceleration marking the transition between Newtonian and deep-MOND regimes, and $\nu(y)$ is the ``interpolating function'' (IF) that connects the two regimes. 

In particular, according to MOND, the underlying baryonic distribution completely governs the total gravitational potential of a galaxy. Since the luminosity profile closely maps the stellar mass density in a galaxy, and by extension, its baryonic distribution, Renzo's rule arises naturally. From the perspective of MOND, Renzo's rule could be formulated more physically for the radial dynamics of late-type galaxies as:

\vspace{1mm}
\emph{``For any feature in a galaxy's total observed RC (\Vobs/), there is a corresponding feature in the RC predicted from the observed baryonic distribution (\Vbar/), and vice versa.''}
\vspace{1mm}

\noindent This provides a simpler, more direct interpretation for testing Renzo's rule with RCs, thus is more suitable for the following study; we will henceforth refer to this statement as Renzo's rule, unless otherwise specified.

Renzo's rule is closely related to, but distinct from, a similar astrophysical phenomenon known as the Radial Acceleration Relation (RAR; \citealt{McGaugh_2016, Lelli_2017, Desmond:2023urj, Varasteanu2025}), which describes a tight local correlation between the gravitational acceleration due to the total mass of a galaxy ($g_{\text{obs}}$) and that due to its baryonic component ($g_{\text{bar}}$). Similar to Renzo's rule, this connection persists in supposedly DM-dominated regions. If the RAR were completely free of intrinsic scatter it would contain Renzo's rule as the only way to produce a monotonic $g_\text{bar}-g_\text{obs}$ relation.
However, the RAR appears to possess a small but non-zero scatter~\citep{McGaugh_2016, Desmond:2023urj, Varasteanu2025}, and its overall shape is dominated by dynamics either towards the outer regions of galaxies with asymptotically flat \Vobs/, or around the innermost parts, where baryons dominate such that $g_{\text{obs}} \approx g_{\text{bar}}$; i.e., the RAR is a ``summary'' statistic of RCs across many galaxies. This can mask the fine-grained details of the $V_\text{bar}-V_\text{obs}$ relation in individual galaxies, as described by Renzo's rule. By focusing on individual galaxies with RCs that exhibit clear features, a systematic test of Renzo's rule can provide important insight into these potentially anomalous connections between DM and galaxy dynamics independently of the RAR.

Although widely acknowledged and cited in the literature, often as supporting evidence for \LCDM/-alternatives like MOND, Renzo’s rule is entirely informal, based entirely on visual inspection of RCs (e.g., see \citealt{Sancisi:2003xt, Gentile_2010, Famaey_2012, Santos_Santos_2015, McGaugh_2019}).
This paper aims to assess the rule's validity through a more systematic study: we examine features in both \Vobs/ and \Vbar/ of each galaxy and develop statistics to quantify their correlation.
Specifically, with a systematic feature identification algorithm, we compute Pearson coefficients and dynamic time warping (DTW) alignment costs on the identified features.
We also develop a set of MOND and $\Lambda$CDM mocks, taking into account prior uncertainties in galaxy properties and observational scatter, to assess the extent to which the observed correlations are compatible with those expected in the two theories, and investigate the properties a data set would need to have for a high-significance detection.

The paper is organized as follows.
Section~\ref{sec:data} introduces the datasets used in our study, including RCs from both observations and simulations.
Section~\ref{sec:methods} then outlines the procedure for assessing the significance of Renzo's rule in each galaxy.
The results are presented in Section~\ref{sec:results}, which includes an investigation on the data quality required for testing Renzo's rule in future galaxy surveys.
Finally, the implications of our results are explored in Sections~\ref{sec:discussion}, and the key conclusions are summarized in Section~\ref{sec:summary}.

\section{Observed and simulated data}\label{sec:data}
For our analysis, we gathered RCs from multiple sources. This includes observed data from (i) two different datasets of the dwarf spiral NGC 1560, (ii) the Spitzer Photometry \& Accurate Rotation Curves (SPARC) dataset, and (iii) Local Irregulars That Trace Luminosity Extremes, The \HI{} Nearby Galaxy Survey (LITTLE THINGS). We also examined galaxies from \LCDM/ hydrodynamical simulations, as selected and simulated by \cite{Santos_Santos_2015}.

\subsection{NGC 1560}
The ``poster child'' of Renzo's rule is the dwarf spiral galaxy NGC 1560, with its RC first studied in \cite{1991MNRAS.249..523B}.
This is a gas-rich, low surface brightness galaxy with an \HI{} mass of $(8.2\pm0.2)\times10^8$~$M_\odot$ \citep{1992Broeils}, a stellar mass of $\sim 5\times10^8$~$M_\odot$ \citep{Gentile_2010} and a dynamical mass of $\sim 1.2\times10^{10}$~$M_\odot$ (assuming a spherical mass distribution, such that $M=V^2R/G$).
Furthermore, it is at a distance of $2.99\pm0.10$~Mpc~\citep{2013Cosmicflows-2}, with an inclination of $82^\circ\pm1^\circ$~\citep{1992Broeils}.

\begin{figure*}
    \centering
    \begin{subfigure}{0.50\textwidth}
        \centering
        \includegraphics[width=\textwidth]{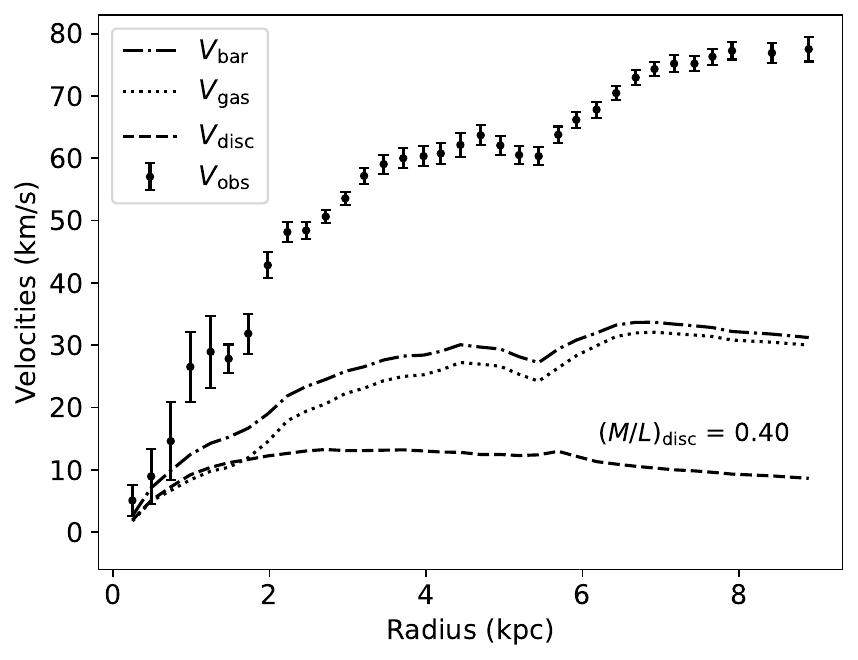}
        \caption{NGC 1560 from digitizing its RCs in \cite{Sanders_2007}, assuming a fixed mass-to-light ratio of $\Upsilon_\text{disc}=0.4$ (B-band) $\pm$ 0.1~dex.}
        \label{subfig:NGC1560_a}
    \end{subfigure}
    \hfill
    \begin{subfigure}{0.48\textwidth}
        \centering
        \includegraphics[width=\textwidth]{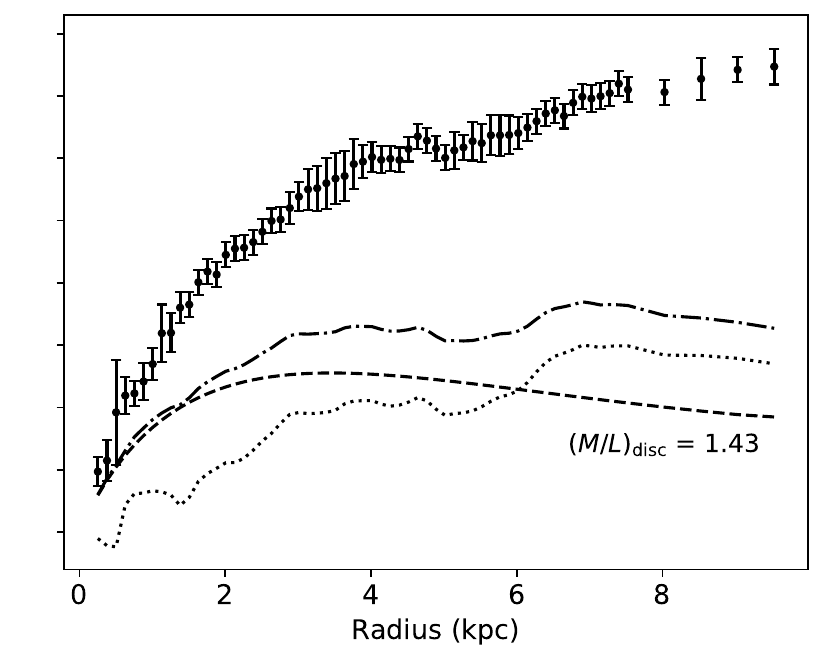}
        \caption{NGC 1560 from \cite{Gentile_2010} (S. McGaugh, priv. comm.), assuming $\Upsilon_\text{disc}=1.43$ (I-band) $\pm$ 0.2 dex.}
        \label{subfig:NGC1560_b}
    \end{subfigure}
    \caption{The two different sets of RCs used for the dwarf spiral galaxy NGC 1560. In both plots, the total observed RC (\Vobs/; error bars), its gaseous component ($V_{\text{gas}}$; dotted line) and its stellar disc component ($V_{\text{disc}}$; dashed line) are given; the total baryonic contribution (\Vbar/; dash-dot line) is calculated using equation~\ref{eqn:Vbar}, assuming uniform mass-to-light ratios of $\Upsilon_{\text{disc}} = 0.4$ and $\Upsilon_\text{disc} = 1.43$, respectively.
    Observe the features at around 4-6~kpc in both dataset, especially the large ``kink'' in Sanders's RCs, which is visible in both \Vobs/ and \Vbar/, thus displaying a clear example of Renzo's rule.}
    \label{fig:NGC1560}
\end{figure*}

The galaxy has a sizeable ``kink'' between 4 and 6~kpc in the gas distribution, which is apparently reflected in the total RC despite the supposed DM domination at that radius (see Fig.~\ref{fig:NGC1560}). This is unnatural under the \LCDM/ model since features in the baryonic RC should be smoothed out in the total RC by the dominant DM halo.

The RC of NGC 1560 has been mapped out several times in the literature ~\citep{1991MNRAS.249..523B, 1992Broeils, Sanders_2007, Gentile_2010}.
The method that seems to yield the strongest support for Renzo's rule is that of \cite{Sanders_2007}, which displays a strong dip in both $V_\text{obs}$ and its gas component ($V_{\text{gas}}$) at 4-6~kpc.
However, we were unable to acquire a copy of this dataset, and hence were forced to manually digitise its plot from \cite{Sanders_2007}\footnote{This was done with \texttt{Plot Digitizer} (\url{https://plotdigitizer.sourceforge.net}), a free program developed by \cite{PlotDigitizer}.}.
The result is shown in Figure~\ref{subfig:NGC1560_a}.

An alternative data reduction pipeline was used in \cite{Gentile_2010}, with a stellar mass model built on I-band data. These data were provided in private communication by Stacy McGaugh.
The RCs are presented in Fig.~\ref{subfig:NGC1560_b}, assuming a stellar disc mass-to-light ratio of $\Upsilon_{\text{disc}} = 1.43\pm0.2$~dex. We also investigate this dataset, to provide a sense of the systematic uncertainty in Renzo-like correlations associated with different data analysis methods.

\subsection{SPARC}
The SPARC database \citep{Lelli_2016} contains neutral atomic hydrogen (\HI) RCs for 175 late-type galaxies, compiled from around three decades of literature. For 56 of the RCs, this is supplemented by H$\alpha$ data at higher spatial resolution in the inner regions.
The dataset covers a broad spectrum of galactic properties, with $K$-band luminosities ranging from $10^7$ $L_\odot$ to $10^{12}$ $L_\odot$, effective radii from 0.3 to 15~kpc, surface brightness from 5 to 5000 $L_\odot$ pc$^{-2}$, and rotation velocities on the flat part of RCs from 20 to 300 km s$^{-1}$.

For each galaxy, the database provides the contributions of gas ($V_{\text{gas}}$), stellar disc ($V_{\text{disc}}$) and bulge ($V_{\text{bul}}$) to the observed RC. It follows that the total baryonic contribution is
\begin{equation}\label{eqn:Vbar}
    V_{\rm bar} = \sqrt{V_{\text{gas}}^2 + V_{\text{disc}}^2 + V_{\text{bul}}^2},
\end{equation}
where the component are rescaled assuming stellar mass-to-light ratios for the gas, disc and bulge of $\Upsilon_{\text{gas}} = 1.0 \pm 0.04$, $\Upsilon_{\text{disc}} = 0.5 \pm 0.125$ and $\Upsilon_{\text{bul}} = 0.7 \pm 0.175$, respectively. The dataset also provides the measured values and uncertainties on the inclination, $3.6\,\mu m$ luminosity, and the observational errors in the total RC (\Vobs/), which are assumed to be completely uncorrelated within each galaxy.

For our analysis, only galaxies with over 20 data points in each component RC are considered to ensure sufficient data for the statistical tests we perform; this reduces our final sample size to 60 galaxies.

\subsection{LITTLE THINGS}

LITTLE THINGS \citep{Hunter_2012} is a multi-wavelength survey of 41 nearby ($\leq 10.3$~Mpc) dwarf galaxies, at an average distance of 3.7~Mpc. All galaxies have an inclination of $i\leq 70^\circ$ such that structures in the discs can be resolved. The dataset covers a large range of dwarf galactic properties, all of which contain atomic hydrogen.
Excluding obviously interacting galaxies, and galaxies with inclinations below $40^\circ$ (where RC extraction can become biased; \citealt{2016MNRAS.462.3628R}), and further assuming that both the stellar and gaseous components are well-represented by smooth exponential discs (thus no features are present in \Vbar/), \cite{Read_2017} derived the total RCs (\Vobs/) for 19 of these galaxies.
We obtained the data for 18 of these galaxies\footnote{We exclude NGC 6822, which is highly extended relative to the primary beam of the Very Large Array used by LITTLE THINGS; special treatment would be required to extract its RCs accurately.} in tabular form from B. Famaey (priv. comm.) for our analysis.

\subsection{MaGICC and CLUES simulated galaxies}
A study by \cite{Santos_Santos_2015} suggests that Renzo's rule emerges in hydrodynamical \LCDM/ simulations. The feedback parameters in these simulations were tuned to reproduce the stellar mass--halo mass relation from abundance matching at one chosen halo mass (specifically, for the irregular galaxy g15784\_Irr), and once they were set, they remained fixed for simulations of haloes with different masses.
Notably, Renzo's rule was apparently recovered across a range of galaxy masses without its explicit addition, nor the incorporation of any closely related phenomena such as the baryonic Tully--Fisher relation~\citep{2000McGaugh-BTFR}. This suggests that, contrary to naive expectations, Renzo's rule may be a result of non-trivial coupling between different mass components as they co-evolve within a \LCDM/ Universe, which is yet to be understood.

More specifically, the authors presented 22 late-type galaxies, 12 of which are generated using the MaGICC smoothed particle hydrodynamical simulations \citep{Stinson_2012}, and the rest with initial conditions from the CLUES project (Constrained Local UniversE Simulations; \citealt{Gottloeber_2010_CLUES, Yepes_2014}).
These galaxies span a wide range in velocity with $52< V_{\text{flat}} <222$ km$\,$s$^{-1}$, where $V_{\text{flat}}$ is the circular velocity of the asymptotically flat region of each RC.
Of the 12 MaGICC galaxies, 5 are similar to the Milky Way (labelled MW), with stellar masses $6.30\times10^9 \, M_\odot \leq M_* \leq 5.67\times10^{10} \, M_\odot$ and halo masses $4.47\times10^{11} \, M_\odot \leq M_{\text{halo}} \leq 1.49\times10^{12} \, M_\odot$, while the rest, labelled Irregular (Irr), have masses $1.32\times10^8 \, M_\odot \leq M_* \leq 1.46\times10^{10} \, M_\odot$ and $5.23\times10^{10} \, M_\odot \leq M_{\text{halo}} \leq 2.82\times10^{11} \, M_\odot$.
Meanwhile, the 10 CLUES galaxies, named C1 to C10, span a mass range of $3.78\times10^8 \, M_\odot \leq M_* \leq 1.45\times10^{10} \, M_\odot$ and $6.44\times10^{10} \, M_\odot \leq M_{\text{halo}} \leq 7.23\times10^{11} \, M_\odot$.

Notably, the authors identified 4 MaGICC galaxies and 5 CLUES galaxies\footnote{The galaxies are g15807 Irr, g15784 Irr, g1536 MW, g5664 MW, C1, C5, C6, C7 (omitted due to merging) and C8.}, for which they state that features in \Vbar/ are reflected in the corresponding \Vobs/, in support of Renzo's rule.
With the exception of galaxy C7, which is dynamically interacting with a very close companion galaxy, we obtained the data from both sets of simulations (21 galaxies in total) by digitizing the plots shown in \cite{Santos_Santos_2015}, once again using \texttt{Plot Digitizer} \citep{PlotDigitizer}.

\section{Methods}\label{sec:methods}
To test Renzo's rule, we examine features in both \Vobs/ and \Vbar/ for each galaxy; if Renzo's rule holds, we expect a strong correlation between these features. This would be contrary to what is (at least naively) predicted by the \LCDM/ model, where the addition of a smooth (e.g., NFW) halo would suppress such features in \Vbar/ and give a smoother \Vobs/.

Our analysis involves four main steps.
First, we generate versions of \Vobs/ as predicted by MOND (\Vmond/) and \LCDM/ (\Vcdm/) given the observed distribution of baryons. Details of this prediction are given in Section~\ref{sec:MONDLCDMRCs}.
Next, to separate out small-scale features from the smooth, larger-scale shape of the RC, residuals are calculated by applying a Gaussian process regression (GPR) to each RC (Section~\ref{sec:residuals}).
Then, a simple feature identification algorithm is used on these residuals to detect galaxies exhibiting features in either or both \Vobs/ and \Vbar/ (Section~\ref{sec:ft_id}).
Finally, correlation statistics, including Pearson correlation coefficients and dynamic time warping, are computed on these residuals, and the results from the data are compared to those with \Vobs/ replaced by \Vmond/ or \Vcdm/ (Section~\ref{sec:correlations}).

\subsection{\texorpdfstring{MOND and \LCDM/ rotation curves}{MOND and LCDM rotation curves}}\label{sec:MONDLCDMRCs}

We employ two different methods to generate \Vmond/ and \Vcdm/.
The first involves Markov-Chain Monte Carlo (MCMC) fits of \Vobs/, treating the galaxy properties and DM halo parameters as free parameters, with priors based on observational data. The second utilizes direct Monte Carlo (MC) sampling on the same priors to generate many instances of \Vbar/, which are used to create independent samples of \Vmond/ and \Vcdm/; these model, respectively, the expected shapes of \Vobs/ (with uncertainties) under a MOND and \LCDM/ universe.
As will be explained below, this corresponds to the difference between tuning the free parameters to match the data best (``MCMC method'') versus simply propagating the prior distributions to create a priori expectations for the correlation statistics (``MC method'').

In both approaches, \Vmond/ follows Milgrom's formula (equation~\ref{eqn:MOND}), where we adopt the ``simple'' interpolating function \citep{Famaey_2005},
\begin{equation}\label{eqn:simpleIF}
    \nu(y) = \frac{1+(1+4y^{-1})^{1/2}}{2},
\end{equation}
which is found to provide an excellent fit to SPARC galaxy kinematics (see \citealt{Li_2018, Stiskalek:2023amy, Desmond:2023urj, Desmond:2023wqf, Desmond:2024eic}).

For \Vcdm/, we assume DM haloes are well-modelled by the Navarro–Frenk–White profile (NFW;~\citealt{Navarro_1996})
\begin{equation}\label{eqn:nfw}
    \rho_{\mathrm{NFW}}(r)=\frac{\rho_{s}}{\left(\frac{r}{r_{s}}\right)\left[1+\left(\frac{r}{r_{s}}\right)\right]^{2}},
\end{equation}
where $r_s$ is a scale radius and $\rho_{s}$ a characteristic density, both of which are free parameters to be determined by MCMC fits or subhalo abundance matching, as will be explained below.

\subsubsection{MCMC fits}\label{sec:MCMC_fits}
For the first method, i.e., fitting of \Vmond/ and \Vcdm/ with MCMC, we imposed truncated Gaussian priors on the distance, inclination, and luminosity of each galaxy.
The best-fit values from prior inferences are used as the means, and their corresponding uncertainties are taken as the standard deviations; the distance and luminosity are required to be positive, while the inclination is required to lie between 15$^\circ$--150$^\circ$. Truncated Gaussian priors are also imposed on the mass-to-light ratios, with means and widths determined by the fiducial value(s) $\Upsilon_{\text{disc}} = 0.5 \pm 0.125$ and, if a bulge exists, $\Upsilon_{\text{bul}} = 0.7 \pm 0.175$; both ratios are required to be positive.

For \Vcdm/ (NFW halo fits), we impose uniform priors $(5.0, 15.0)$ on the logarithmic\footnote{In this paper, $\log$ denotes base-10 logarithm, whereas natural logarithms are denoted by $\ln$.} mass, $\log M_{200}$, where~$M_{200}$ is defined as the mass contained within a sphere of radius $r_{200}$, within which the average density is 200 times the critical density of the Universe, i.e.
\begin{equation}
    M_{200} = \frac{4}{3}\pi r_{200}^3 \times 200\,\rho_\text{crit} = \frac{4}{3}\pi r_{200}^3 \times 200 \left(\frac{3H^2}{8\pi G}\right).
\end{equation}
We also impose Gaussian priors $\mathcal{N}(\log c_{200}, 0.11)$ on the logarithmic concentration, $\log c$, where $c$ is defined as the ratio of the virial radius to the scale radius, $c = r_{200}/r_s$, and $\log c_{200}$ satisfies the concentration-mass relation:
\begin{equation}
    \log c_{200} = 0.905 - 0.101 \log (M_{200}/[10^{12}h^{-1}M_\odot]).
\end{equation}
This relation is well-described by a log-normal distribution with an intrinsic scatter of 0.11 dex~\citep{10.1111/j.1365-2966.2008.14029.x,Dutton_2014}. By integrating equation~\ref{eqn:nfw} and relating $r_s$ to $r_{200}$ via $r_s = r_{200}/c$, the DM component of \Vcdm/ is derived to be
\begin{equation}
    V_{\text{DM}}^2 (r) = \frac{GM_{200}}{r} \times \frac{f(r/r_s)}{f(c)},
\end{equation}
where we define the function
\begin{equation}
    f(x) \equiv \ln(1+x) - \frac{x}{1+x}.
\end{equation}
In the MOND fits, the exact value of $a_0$ has negligible effect on our analysis (and similar analysis of baryon--DM dynamics in general, e.g., see \citealt{Li_2018}), and is therefore treated as a fundamental constant with value $a_0=1.20 \times 10^{-10}$~m~s$^{-2}$; thus no additional parameters are fitted for MOND.

For both profiles, we obtain the fits and their corresponding \Vbar/ --- representing a reasonable realisation of the galaxy's systematic uncertainties --- by drawing 1000 random samples from the posterior using the No U-Turns Sampler (NUTS; \citealt{NUTS_2014}), as implemented in the \texttt{NumPyro} package~\citep{Phan_2019}. To provide a meaningful comparison with \Vobs/, the fits are further scattered by observational uncertainties, which we assume to be Gaussian and completely uncorrelated (as is assumed for SPARC galaxies).

\subsubsection{MC sampling}\label{sec:MC_samp}
For the second method, we generate 1000 instances of \Vbar/ for each galaxy, sampling directly from (truncated) Gaussian uncertainties in the observed galaxy distance, luminosity and mass-to-light ratios. For MOND, these uncertainties are propagated to \Vobs/ by applying Milgrom's formula (Equations~\ref{eqn:MOND} and \ref{eqn:simpleIF}) to each realization. Again, $a_0$ is fixed.

For \LCDM/, we obtained $V_{\text{DM}}$ from the technique of subhalo abundance matching (SHAM), as outlined in \cite{Stiskalek_2021}, which is based on earlier results of abundance matching \citep{Kravtsov_2004, 2004MNRAS.353..189V, Conroy_2006, Behroozi_2010, 2010ApJ...710..903M, Reddick_2013, Lehmann_2016}.
Fundamentally, SHAM postulates the existence of a near-monotonic relation between a galaxy property (e.g., stellar mass or luminosity) and a halo property (often some function of virial mass and concentration).
Assuming the NFW profile (equation~\ref{eqn:nfw}) applies, SHAM provides a method to generate $V_{\text{DM}}$ based solely on the stellar mass of each galaxy, independent of \Vobs/.
The $V_{\text{DM}}$ obtained is then added to each Monte-Carlo realization of \Vbar/ in quadrature, followed by additional Gaussian scattering based on observational uncertainties in \Vobs/ to give \Vcdm/.

Specifically, the SHAM model takes in two free parameters, a generalized halo proxy $m_\alpha$ and a Gaussian scatter $\sigma_\text{AM}$, where $\alpha$ is defined as an interpolation between the peak virial mass over the history of the halo, $M_\text{peak}$, and the present-day value $M_0$:
\begin{equation}
    m_\alpha = M_0 \left( \frac{M_\text{peak}}{M_0} \right)^\alpha .
\end{equation}
For our analysis, we have chosen $\alpha=1.20$ and $\sigma_\text{AM}=0.27$, which provide the best fit for $M_*$-based SHAM \citealt{Stiskalek_2021}).

\subsection{Calculation of residuals}\label{sec:residuals}
While the general profile of each RC is largely governed by the observed mass discrepancy, i.e., the various extent of DM dominance at large radii, Renzo's rule pertains to the small-scale bumps and wiggles distinct from the RC's overall shape.
To extract these features from the otherwise smooth RCs, we turn to Gaussian Process Regression (GPR). 
By using appropriate hyperparameters, the residuals from each GPR, denoted $\delta V_X$ (e.g., $\delta$\Vobs/ denotes the residuals from a GPR on \Vobs/), will contain all, and only, the small-scale features needed for probing Renzo's rule.

Specifically, we fit each RC using a GPR with a squared exponential kernel \citep{10.7551/mitpress/3206.001.0001}, defined as
\begin{equation}
    k_{SE}(x_i, x_j) = \sigma^2\exp[-\frac{(x_i-x_j)^2}{2l^2}] + \epsilon^2_n \delta_{ij},
\end{equation}
where $\sigma^2$ is the variance, $\epsilon^2_n$ is a diagonal noise term, and $l$ is the characteristic length-scale of the kernel; physically, $l$ determines the typical length of `wiggles' allowed by the GPR fit, which should ideally be larger than the width of all visible small-scale features.

In particular, for each galaxy, we assign a common length-scale $l$ to all its RCs (\Vbar/, \Vobs/, \Vmond/, \Vcdm/), ensuring that the overall shape of each RC is captured without overfitting into individual features.
In general, $l$ is chosen to be around half the maximum radius of each galaxy, such that visually, the GPR captures the overall $\arctan$-like shape of \Vobs/.
To account for small amounts of noise and variability, log-normal priors $\ln \mathcal{N}(0,1)$ are also placed on $\sigma^2$ and $\epsilon^2_n$.

\subsection{Feature identification}\label{sec:ft_id}
Features in the residuals are identified systematically using a simple algorithm. Given an RC, we first divide it into segments of positive and negative residuals, and select those containing at least 3 data points with height-to-noise ratios above 2.0, reducing the risk of picking out random fluctuations due solely to observational errors.
For \Vobs/, the noise ($\epsilon_{\text{obs}}$) comes from observational errors, which we assume to be completely uncorrelated, whereas the noise in \Vbar/ ($\epsilon_{\text{bar}}$) is obtained from Monte Carlo sampling on Gaussian uncertainties in the galaxy's distance, luminosity, and mass-to-light ratios. Then, any adjacent segments are merged into a single, larger feature.
By picking out the left and right base of each feature, the position and base width of each feature are also obtained automatically. This is detailed in Algorithm~\ref{alg:ft_id}.

\begin{algorithm}
\caption{Feature identification}\label{alg:ft_id}
\begin{algorithmic}
\State \textbf{Inputs:} GPR residuals ($\delta$\Vobs/, $\delta$\Vbar/) and noise ($\epsilon_{\text{obs}}, \epsilon_{\text{bar}}$)
\\
\State $N \gets$ residuals / noise   \Comment{Normalize residuals by noise}
\State $S_N\gets$ Split($N$)    \\
\Comment{$N$ is split into positive and negative sublists, e.g., $N=[-1, 1, 2, 3, -1, -2] \mapsto S_N=[[-1], [1,2,3], [-1,-2]]$}
\\
\For{each segment $s$ in $S_N$}
\If{$s$ contains 3 or more elements with norm $> 2.0$}
    \State add $s$ to list of features
\EndIf
\EndFor
\\
\For{$f_i$ in list of features}
\If{right base of $f_i=$ left base of $f_{i+1}$}
\State Combine $f_i$ and $f_{i+1}$ into one bigger feature
\EndIf
\EndFor
\\
\State \textbf{Returns:} Left bases, right bases and widths of features
\end{algorithmic}
\end{algorithm}

Specifically, this algorithm is used in two different ways. Firstly, by requiring at least one feature to appear in either \Vobs/ or \Vbar/, it serves as a filter to select galaxies on which Renzo's rule can be meaningfully tested. Secondly, this can be used to improve the residual calculations: the GPR can occasionally overfit the features, such that they are not captured entirely in the residuals; in this case, the algorithm (combined with visual verification) can be used to flag and remove such features from the RC, allowing us to fit a second, less biased GPR.
Note that in principle, this may be performed iteratively, but we find minimal change after the first iteration in all cases.

\subsection{Correlation statistics}\label{sec:correlations}
For each galaxy, we compare the residuals of \Vobs/ ($\delta$\Vobs/) and those of \Vbar/ ($\delta$\Vbar/) using two correlation statistics: Pearson correlation coefficient and dynamic time warping (DTW) cost.
The same statistics are also computed between $\delta$\Vbar/ and the corresponding $\delta$\Vmond/ and $\delta$\Vcdm/.

It is worth mentioning that although the Spearman coefficient is often preferred for investigating non-linear correlations, it is not used here because it correlates ranked values instead of raw magnitudes.
Specifically, Renzo's rule deviates from \LCDM/ expectations in DM-dominated regions, where features in \Vbar/ should be largely, but not entirely, washed out by the dominant contribution of $V_{\text{DM}}$ when summing in quadrature to form \Vobs/ (see discussion in Section~\ref{sec:LCDM_Renzo}).
On the other hand, Renzo's rule posits that the feature sizes in \Vobs/ and \Vbar/ remain relatively similar (as predicted, for instance, by MOND).
Distinguishing these two scenarios is crucial to our analysis, which requires preserving actual values rather than only ranks.

\subsubsection{Pearson correlation}
The Pearson correlation coefficient \citep{doi:10.1098/rspl.1895.0041} is defined as
\begin{equation}
    \rho_p
    \equiv \frac{\sum(x_i - \bar{x})(y_i-\bar{y})}{\sqrt{\sum(y_i-\bar{y})^2 \sum(x_i-\bar{x})^2}}
\end{equation}
for two arrays $\{x_i\}_{i=0}^n, \{y_i\}_{i=0}^n$ with means $\bar{x}, \bar{y}$, respectively.
$\rho_p$ ranges from -1 to 1, with 1 representing perfect linear correlation between the two arrays, -1 representing perfect linear anti-correlation, and 0 indicating an absence of linear correlation.

To better understand its behaviour as a function of radial distance, we calculate $\rho_p$ from data in spheres of increasing radii, starting from the radius that encloses 5 data points and ending at the final measured data point.

\subsubsection{Dynamic time warping (DTW)}\label{sec:dtw}
Originally developed for speech applications \citep{Vintsyuk1968SpeechDB}, DTW is an efficient algorithm for measuring the similarity between two series of data.
Given two arrays, $\textbf{A} = \{A_i\}_{i=1}^m$ and $\textbf{B} = \{B_j\}_{j=1}^n$ (in our case, \Vobs/ and \Vbar/, so $m=n$),  DTW matches their elements following four rules:
\begin{enumerate}
    \item Every point in $\{A_i\}$ is matched with at least one point in $\{B_j\}$, and vice versa;
    \item $A_1$ is matched with $B_1$ (beginnings must match);
    \item $A_m$ is matched with $B_n$ (ends must match);
    \item The matching of indices, when viewed as a function $j\equiv\pi(i)$, forms a monotonically increasing sequence, i.e., if $i_1 < i_2$ and $A_{i_1}$ (resp. $A_{i_2}$) is matched with $B_{j_1}$ ($B_{j_2}$) such that $j_1=\pi(i_1)$ and $j_2=\pi(i_2)$, then $j_1 \leq j_2$.
    \item Assuming features are locally correlated (e.g., a feature at the start of \textbf{A} can only influence data points near the start of \textbf{B}), which is likely the case for features in RCs, we only consider matches of indices within a window of $\pm5$ data points (inclusive), i.e., $\pi(i) \in [i-5, i+5]$.
\end{enumerate}
We define the optimal alignment as the minimum Euclidean separation
\begin{equation}\label{eqn:dtw}
    \text{DTW}(\textbf{A},\textbf{B}) = \min_{\pi\in\mathcal{M}} \left\{\sum_{i=1}^m |A_i - B_{\pi(i)}|\right\},
\end{equation}
where $\mathcal{M}$ is the set of all valid matches. DTW seeks this optimal alignment efficiently using dynamic programming.
In other words, DTW is a flexible algorithm that works by shifting, inserting and/or merging elements in the two respective series under reasonable restrictions, in order to identify the alignment which minimizes their Euclidean separation.

\begin{figure}
\centering
    \includegraphics[width=\linewidth]{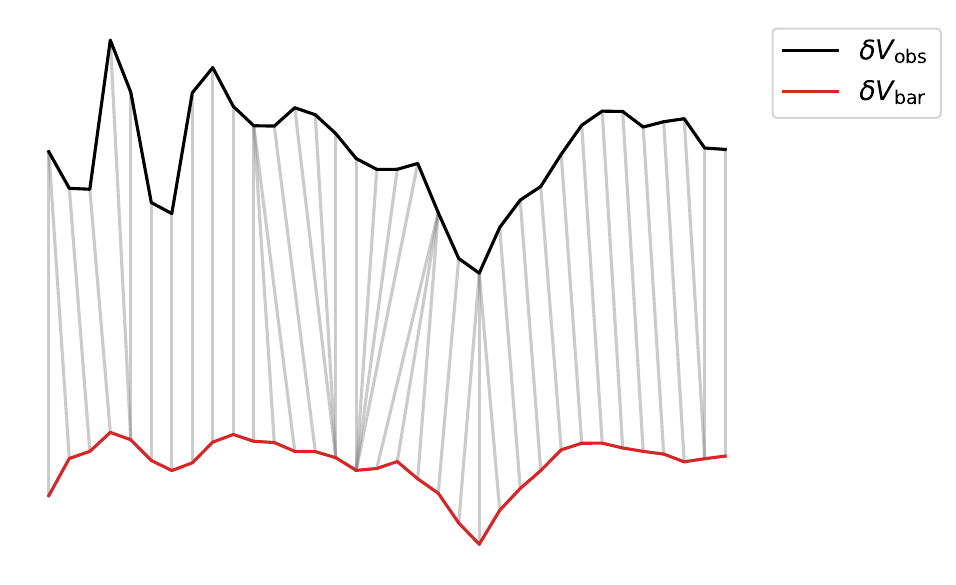}
    \caption{We demonstrate how DTW aligns $\delta$\Vobs/ to $\delta$\Vbar/ in Sanders's NGC 1560 data.
    Here, the grey lines depict the optimal mapping $\pi\in\mathcal{M}$ which minimizes the Euclidean separation of $\delta$\Vobs/ and $\delta$\Vbar/ to give the (normalized) DTW alignment cost, DTW$(\delta$\Vobs/$, \delta$\Vbar/$)$, as defined in equation~\ref{eqn:dtw}.
    Note that the two sets of residuals are shifted apart vertically to aid visualization.}
    \label{fig:DTW_example}
\end{figure}

In our analysis, DTW is used to compare $\delta$\Vobs/, $\delta$\Vmond/ and $\delta$\Vcdm/ to $\delta$\Vbar/ (e.g., see Fig.~\ref{fig:DTW_example}). Specifically, the (normalized) alignment cost for each RC, $V_X$, with residuals $\delta V_X$, is defined as DTW($\delta V_X,$ $\delta$\Vbar/) via equation~\ref{eqn:dtw}, divided by the total number of data points in the two curves. Thus a cost of 0 indicates the residuals of a RC is completely identical to $\delta$\Vbar/, while higher costs reflect decreasing similarity.

\section{Results}\label{sec:results}
\subsection{NGC 1560}\label{sec:NGC1560_results}
Our algorithm identified a feature around 4.2-6.2~kpc in both \Vbar/ and \Vobs/ of Sanders's NGC 1560 (from manual digitization). Specifically, both features display a significant signal-to-noise ratio (above 2.0) around the feature, triggering the feature detection algorithm. Thus we expect this set of RCs to be useful in testing Renzo's rule.

\begin{figure*}
    \centering
    \begin{subfigure}{0.525\textwidth}
        \centering
        \includegraphics[width=\textwidth]{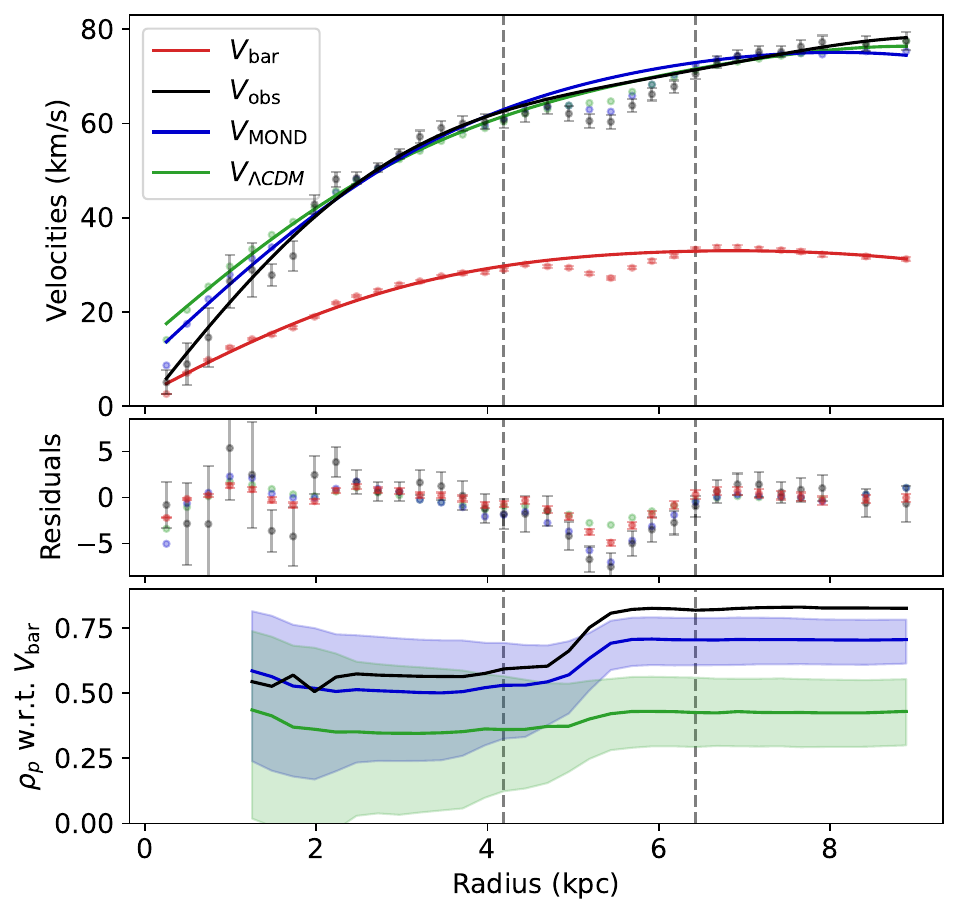}
    \end{subfigure}
    \hfill
    \begin{subfigure}{0.465\textwidth}
        \centering
        \includegraphics[width=\textwidth]{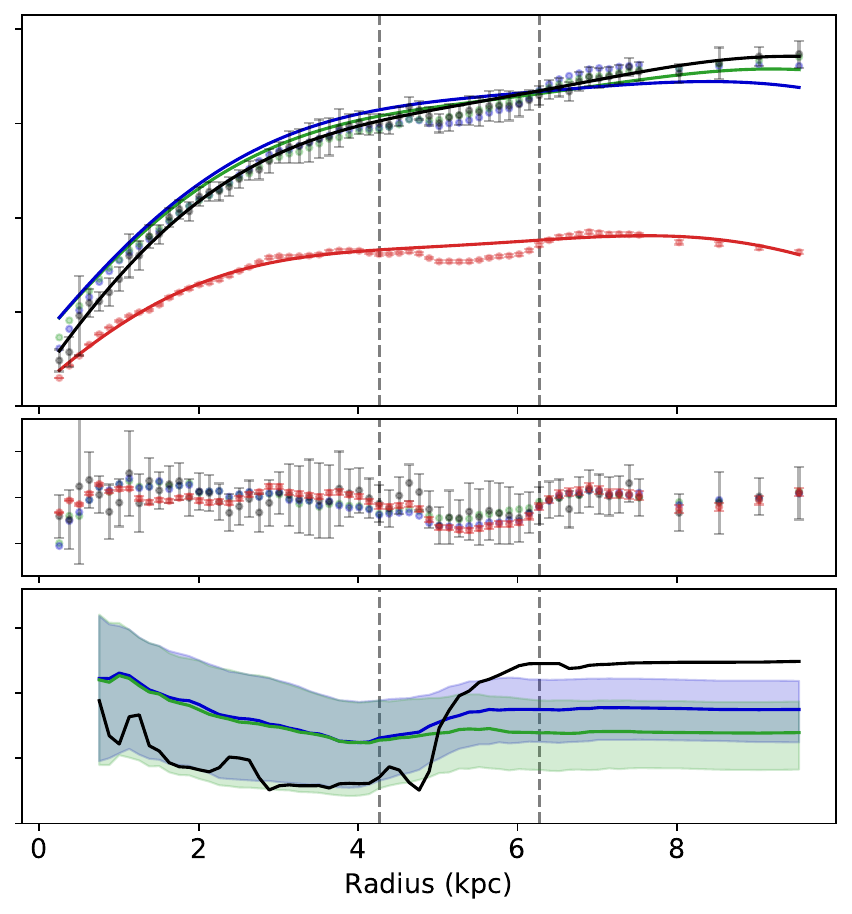}
    \end{subfigure}
        \caption{Statistics of \Vobs/, \Vmond/ and \Vcdm/ relative to \Vbar/ for the two versions of NGC 1560; (\emph{left}) RCs obtained from \protect\cite{Sanders_2007} using \texttt{Plot Digitizer}, and (\emph{right}) RCs from \protect\citealt{Gentile_2010} (S. McGaugh, priv. comm.). \Vmond/ and \Vcdm/ are generated with MCMC fits.
        Each plot contains three panels showing, from top to bottom: the various data and fits, residuals from the GPR (with kernel length-scale $l=4.5$~kpc, and fitted with points within the 4.2-6.2~kpc window removed), and the Pearson coefficients as a function of maximum radii, where the bands outline the $1\sigma$-confidence range around the displayed means (solid lines) as obtained from the 1000 realizations of each RC.
        The dashed vertical lines indicate the boundaries of the 4.2-6.2~kpc window, within which features are identified in both \Vobs/ and \Vbar/ of Sanders's RCs, but not in that of Gentile et al. -- note that in the original GPR, i.e., without removing points in the 4.2-6.2~kpc window, the now-visible dip in Gentile et al. is part of a larger `wiggle' at around 4.5-7.5~kpc, which does not achieve a signal-to-noise ratio of 2.0 to trigger the feature identification algorithm; of course, with the new, refitted GPR, the large dip is identified by Algorithm~\ref{alg:ft_id}, but it is technically not a feature since points are altered for the new GPR (which may have introduced bias in its residuals).
        Error bars for MOND and \LCDM/, which are of the same size as that in \Vobs/, are suppressed to avoid clutter.
        Note that we only show the correlation coefficients at $\geq5$ data points since, before that, $\rho_p$ simply fluctuates unpredictably between -1 and 1 due to pure noise.}
    \label{fig:corr_NGC1560}
\end{figure*}

In this case, we first generated \Vmond/ and \Vcdm/ using MCMC fits (following Section~\ref{sec:MCMC_fits}). Then, a second, less biased GPR is fitted to each curve, where data points on all RC components that lie within the identified 4.2-6.2~kpc window are removed.
With these residuals, we found that the Pearson correlation between $\delta$\Vobs/ and $\delta$\Vbar/, denoted $\rho_p(\delta$\Vobs/, $\delta$\Vbar/), is higher than both $\rho_p(\delta$\Vmond/, $\delta$\Vbar/) and $\rho_p(\delta$\Vcdm/, $\delta$\Vbar/), with slight preference for stronger-correlated MOND statistics.
Plotting the correlation coefficients up to a maximum radius, and varying that radius, it is clear that the strong correlation is largely due to the ``kink'' at 4.2-6.2~kpc (as shown in Fig.~\ref{fig:corr_NGC1560}).
However, with DTW, the statistics for \LCDM/ and MOND are very similar, with the observed data compatible with both sets of predictions (see Fig.~\ref{fig:DTW_example} and first row of Table~\ref{tab:DTW1560}).

\begin{table}
    \centering
    \begin{tabular}{c|c|c|c}
        \hline
         & \Vobs/ & \Vmond/ & \Vcdm/ \\
        \hline \\[-1em]
         \cite{Sanders_2007} & 0.62 & $0.71^{+0.16}_{-0.13}$ & $0.70^{+0.14}_{-0.13}$ \\[0.5em]
         \cite{Gentile_2010} & 0.30 & $0.99^{+0.13}_{-0.11}$ & $0.95^{+0.12}_{-0.11}$ \\[0.1em]
         \hline
    \end{tabular}
    \caption{DTW costs of residuals relative to $\delta$\Vbar/ for the two different versions of NGC 1560, with $1\sigma$ uncertainties derived from the 1000 samples in our MCMC fitting procedure.
    Note that, as opposed to Pearson coefficients, a lower DTW costs indicates a stronger correlation.
    We see that in both cases, DTW fails to provide a clear separation between the expectations from MOND and \LCDM/, but indicates an exceedingly strong correlation in the data (especially for that from Gentile et al.), even beyond that expected by MOND.}
    \label{tab:DTW1560}
\end{table}

For our second version of NGC 1560 (priv. comm., S. McGaugh; data from \citealt{Gentile_2010}), no features were identified in both \Vobs/ and \Vbar/. Motivated by our first dataset, we once again generated MOND and \LCDM/ analogues and refitted GPRs, where points within the 4.2-6.2~kpc window are removed to minimize bias towards the reputed feature. Indeed, with these residuals, we do find a large dip in \Vbar/, which our algorithm does not originally identify as a feature since it was `over-fitted' by the GPR (thus is technically not a feature as defined in Section~\ref{sec:ft_id}).
For this dataset, both the Pearson coefficients and DTW costs indicate an exceptionally strong correlation between \Vobs/ and \Vbar/, stronger than what is naively expected from the similar statistics of MOND and \LCDM/ (this is especially true for DTW statistics; see second row of Table~\ref{tab:DTW1560}).

\begin{table}
    \centering
    \begin{tabular}{c|c|c|c|c}
        \hline
        Deviation (in $\sigma$) w.r.t. & \multicolumn{2}{c}{MOND} & \multicolumn{2}{c}{\LCDM/} \\
        \hline
         & $\rho_p$ & DTW & $\rho_p$ & DTW \\[0.1em]
         Sanders & 1.53 & -0.62 & 3.2 & -0.59 \\
         Gentile et al. & 1.5 & -5.75 & 2.08 & -5.65 \\
         Sanders (window) & 1.36 & 0.05 & 1.68 & -0.09 \\
         Gentile et al. (window) & 2.5 & -2.94 & 3.06 & -3.1 \\
         \hline
    \end{tabular}
    \caption{We summarize the normalized deviation (in $\sigma$) of the correlation statistics obtained from both datasets of NGC 1560, relative to expectations from MOND and \LCDM/ (NFW haloes).
    Note that a positive deviation in Pearson coefficients (denoted $\rho_p$) indicates stronger correlation than predicted, whereas the signs are swapped for DTW, i.e., stronger correlations are reflected as negative deviations.
    In most cases, we find an exceedingly strong correlation for features in the data, even beyond the corresponding expectations from MOND (with up to $5\sigma$ deviation in the second case).
    These statistics likely reflect the overestimation and/or strong correlation of errors in the data, rather than a definitive preference or rejection of either theory.}
    \label{tab:NGC1560_summary}
\end{table}

\begin{figure*}
    \centering
    \includegraphics[width=\textwidth]{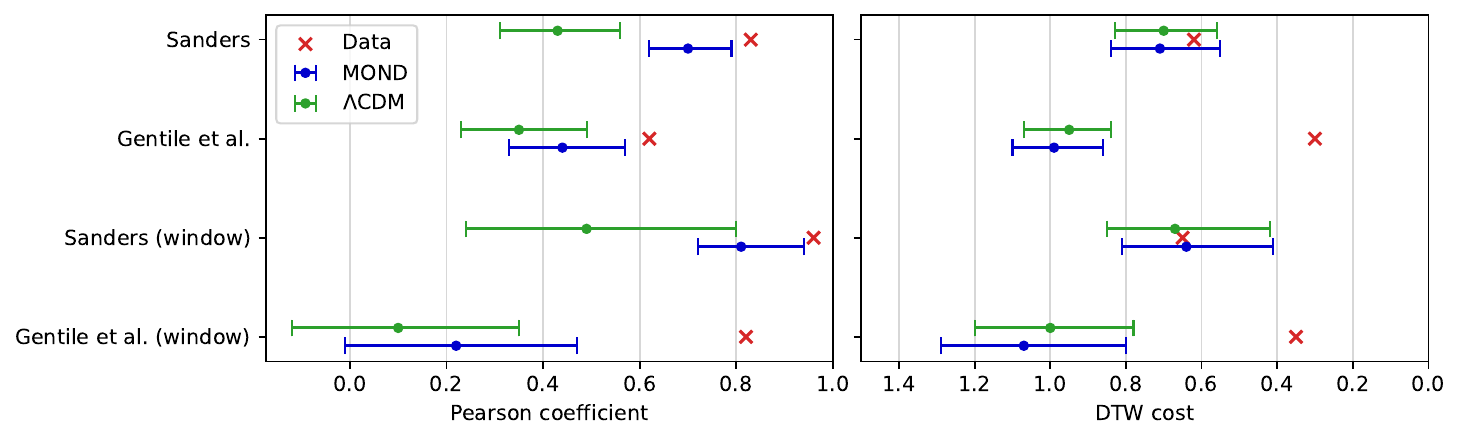}
    \caption{A summary of all correlation statistics on NGC 1560 --- Pearson coefficient (left) and DTW cost (right; here the x-axis is reversed such that stronger correlations appear towards the right in both plots). The four datasets involved, ordered from top to bottom, are (i) the complete RCs in \protect\cite{Sanders_2007} from plot digitization, (ii) the complete RCs in \protect\cite{Gentile_2010} from priv. comm. with S. McGaugh, (iii) Sanders's RCs restricted to 4.2-6.2~kpc, where a feature is identified in both \Vbar/ and \Vobs/, and (iv) Gentile's RCs, similarly restricted to 4.2-6.2~kpc, but where no features were identified.
    Our results indicate that in most cases, especially in the RCs from Gentile et al., features in \Vobs/ and \Vbar/ of correlate stronger than what is expected from \LCDM/ (specifically that from a smooth NFW halo) and even MOND.
    However, in many cases (especially for DTW), the same statistics fail to separate clearly the expectations from MOND and \LCDM/, likely due to large and strongly correlated uncertainties the data, which over-scatter the corresponding Monte-Carlo samples, weakening the expected correlations.
    This suggests that a more detailed study of the dwarf spiral is required to better understand the origin and strong correlations of its features.}
    \label{fig:NGC1560_summary}
\end{figure*}

Using the information from our algorithm, we repeated the analysis on both datasets, this time restricted to the radial window of 4.2-6.2~kpc.
The results, along with statistics from the previous `full analyses', are shown in Fig.~\ref{fig:NGC1560_summary}.
These results are very similar to what was obtained in the full analyses. Notably, across all datasets, the Pearson coefficients indicate stronger correlation in MOND than \LCDM/ as one naively expects, whereas with the many overlapping error bars, DTW appears to be inadequate in differentiating between the two predictions. This suggests that Pearson coefficients may be better suited than DTW for testing Renzo's rule; a systematic comparison between Pearson coefficients and DTW, using mock data, will be presented in Section~\ref{sec:mock_data}.

By taking the average uncertainties on each error bar in Fig.~\ref{fig:NGC1560_summary}, we can also estimate the significance of deviation for the observed statistics relative to that predicted by MOND and \LCDM/, which are presented in Table~\ref{tab:NGC1560_summary}.
If the observational errors in \Vobs/ are accurate and completely uncorrelated, these results would indicate a surprisingly strong correlation in both datasets, exceeding even what is expected by MOND by upwards of $5\sigma$.
But, as hinted, this unlikely result is possibly explained by (i) overestimated errors in \Vobs/; and/or (ii) a correlation between the data points, such that the correlation statistics are in fact underestimated by our generated MOND and \LCDM/ samples, where we assumed no covariance.
Indeed, both caveats are especially crucial for our analyses on the densely sampled RCs from Gentile et al., which contains visibly larger and more correlated uncertainties than that in Sanders's dataset.

Nevertheless, assuming such overestimations and/or correlations are not large (as is likely the case for Sanders's dataset), these results suggest that fluctuations in \Vobs/ tend to agree more with expectations from MOND than \LCDM/ (specifically that from a NFW halo fit), supporting earlier claims of Renzo's rule by \cite{Sanders_2007} and \cite{Gentile_2010}.
That said, more detailed studies and modelling of the dwarf spiral is evidently required to better understand the origin and strong correlation of its features, before any conclusive claims can be established.

\subsection{SPARC galaxies}
\begin{figure*}
    \centering
    \includegraphics[width=\linewidth]{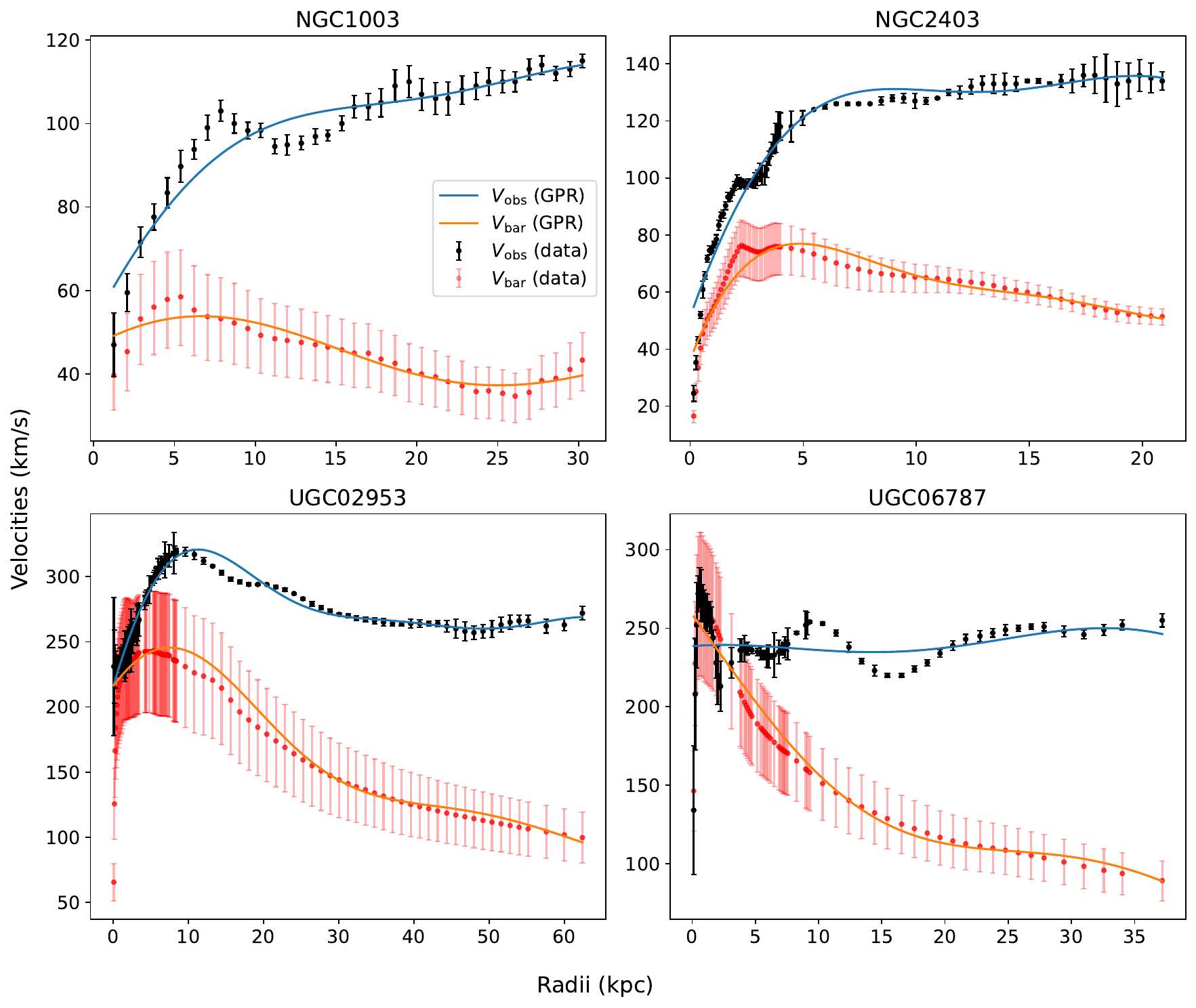}
    \caption{We display four SPARC galaxies where at least one feature is identified in \Vobs/ (black error bars), but not in \Vbar/ (red error bars). The blue and orange solid lines show the GPR fits for \Vobs/ and \Vbar/ respectively, relative to which features are identified.
    While there are interesting cases where prominent features are visible in \Vobs/ without similar fluctuations in \Vbar/, most notably in UGC 6787 (as shown on the bottom-right panel), strongly correlated (and possibly overestimated) a priori uncertainties in galaxy properties such as distances, mass-to-light ratios, etc., also contributed heavily to the lack of identifiable features in \Vbar/; these uncertainties are treated as uncorrelated by Algorithm~\ref{alg:ft_id}, making it practically impossible for fluctuations in \Vbar/ to achieve the threshold signal-to-noise ratio of $T=2.0$.}
    \label{fig:SPARC_examples}
\end{figure*}

Out of the 60 SPARC galaxies, a total of 31 features are identified in \Vobs/ in the RCs of 25 galaxies, whereas no features are identified in any of the \Vbar/ measurements.
Note that while there are a few interesting cases where prominent features are visible in \Vobs/ without similar fluctuations in \Vbar/, strongly correlated (and possibly overestimated) a priori uncertainties in galaxy properties such as distances, mass-to-light ratios, etc., also contributed heavily to the lack of identifiable features in \Vbar/; these uncertainties are treated as uncorrelated by Algorithm~\ref{alg:ft_id}, which makes it practically impossible for fluctuations in \Vbar/ to achieve the threshold signal-to-noise ratio of $T=2.0$ (see Fig.~\ref{fig:SPARC_examples}).
Nevertheless, \Vmond/ and \Vcdm/ are generated using Monte-Carlo sampling (as outlined in Section~\ref{sec:MC_samp}), and the same procedures of residual calculations with GPR and correlation calculations are applied across each set of RCs.

\begin{figure*}
    \centering
    \begin{subfigure}{\textwidth}
        \centering
        \includegraphics[width=\textwidth]{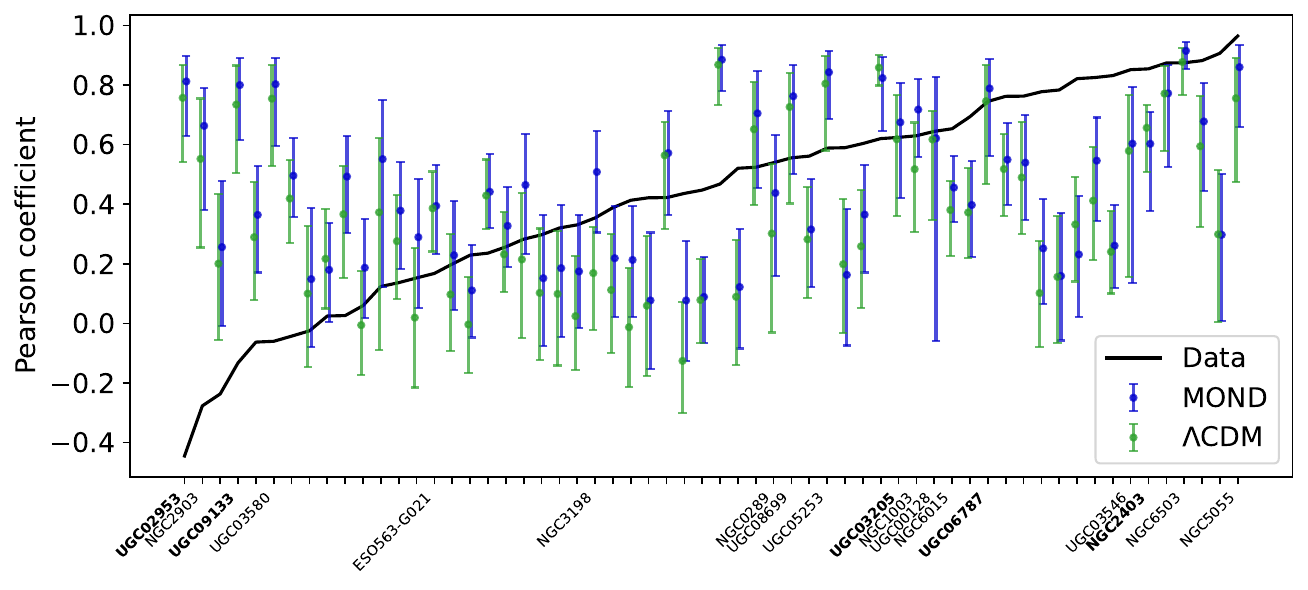}
    \end{subfigure}
    \hfill
    \begin{subfigure}{\textwidth}
        \centering
        \includegraphics[width=\textwidth]{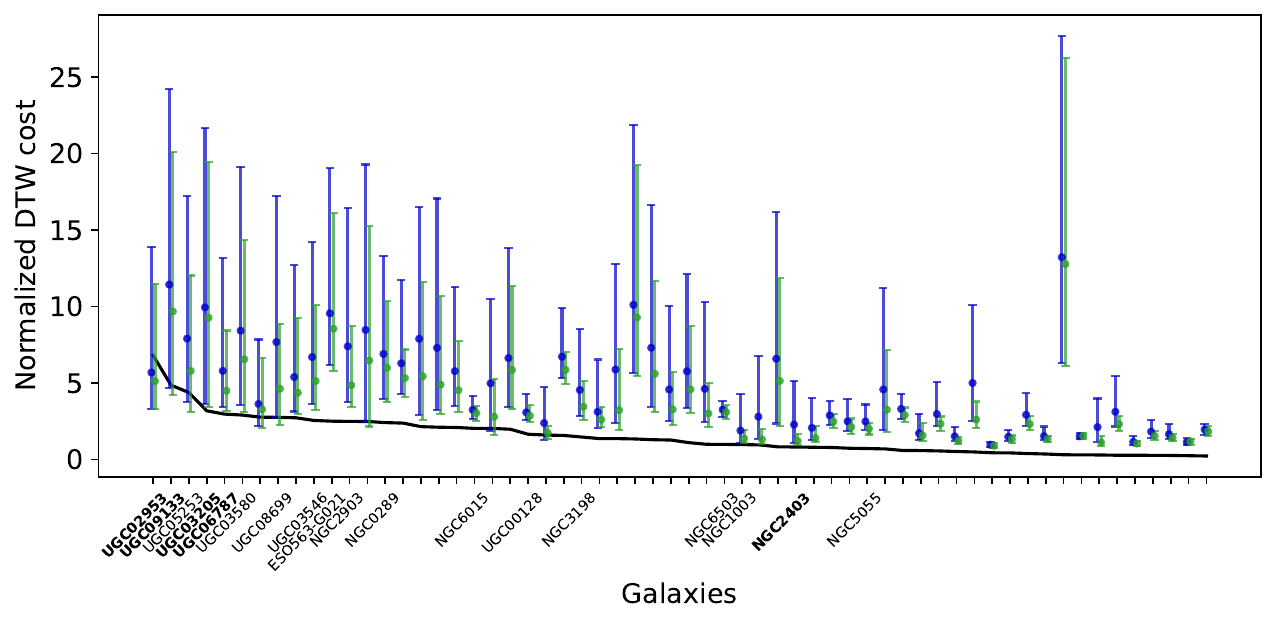}
    \end{subfigure}
        \caption{Statistics of Pearson coefficients (top) and DTW (bottom) on SPARC data. In both plots, the black solid line outlines the statistics between $\delta$\Vobs/ and $\delta$\Vbar/ in ascending correlations, i.e. ascending Pearson coefficients and descending DTW costs. The blue and green error bars show the $1\sigma$ statistics (from MC samples) of $\delta$\Vmond/ and $\delta$\Vcdm/, respectively, with respect to $\delta$\Vbar/. Galaxies labelled along the x-axis contain feature(s) in \Vobs/ (but not \Vbar/), as identified by the feature identification algorithm (see sec~\ref{sec:ft_id}); those that are in bold contain 2 features in \Vobs/, while those in regular font contain only one.
        We see that in most cases, likely due to the high level of noise and lack of features in \Vbar/, both the Pearson coefficient and DTW cost fail to discern between the statistics expected from MOND and \LCDM/, thus no clear conclusion can be made on the validity of Renzo's rule.
        }
        \label{fig:SPARC_hist}
\end{figure*}

A summary of the correlation statistics for all SPARC galaxies is presented in Fig.~\ref{fig:SPARC_hist}, where galaxies are arranged in ascending correlations between $\delta$\Vobs/ and $\delta$\Vbar/, i.e. ascending Pearson coefficients and descending DTW costs. Note that in most cases, due to the high level of noise and lack of features in \Vbar/, both the Pearson coefficient and DTW cost fail to discern between the statistics expected from MOND and \LCDM/. Interestingly, galaxies with a feature(s) identified in \Vobs/ (but not \Vbar/) tend to have higher DTW costs, whereas there is not a similar pattern in the Pearson coefficient distribution.

\begin{figure*}
    \centering
    \begin{subfigure}{\textwidth}
        \centering
        \includegraphics[width=\textwidth]{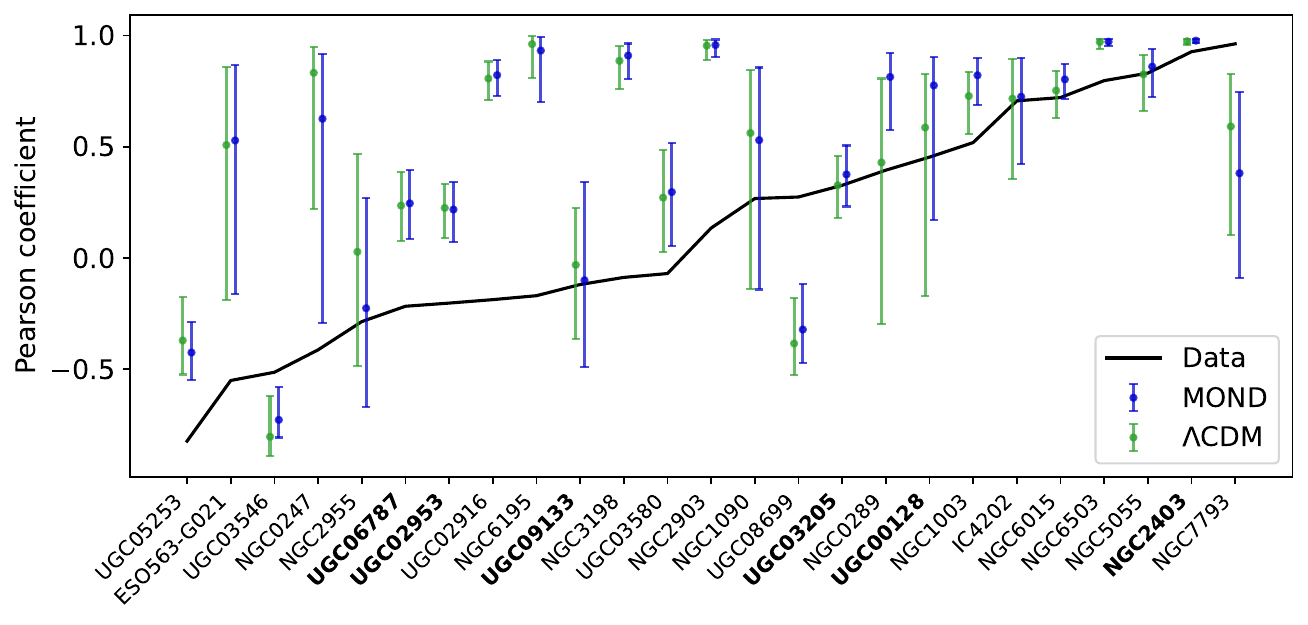}
    \end{subfigure}
    \hfill
    \begin{subfigure}{\textwidth}
        \centering
        \includegraphics[width=\textwidth]{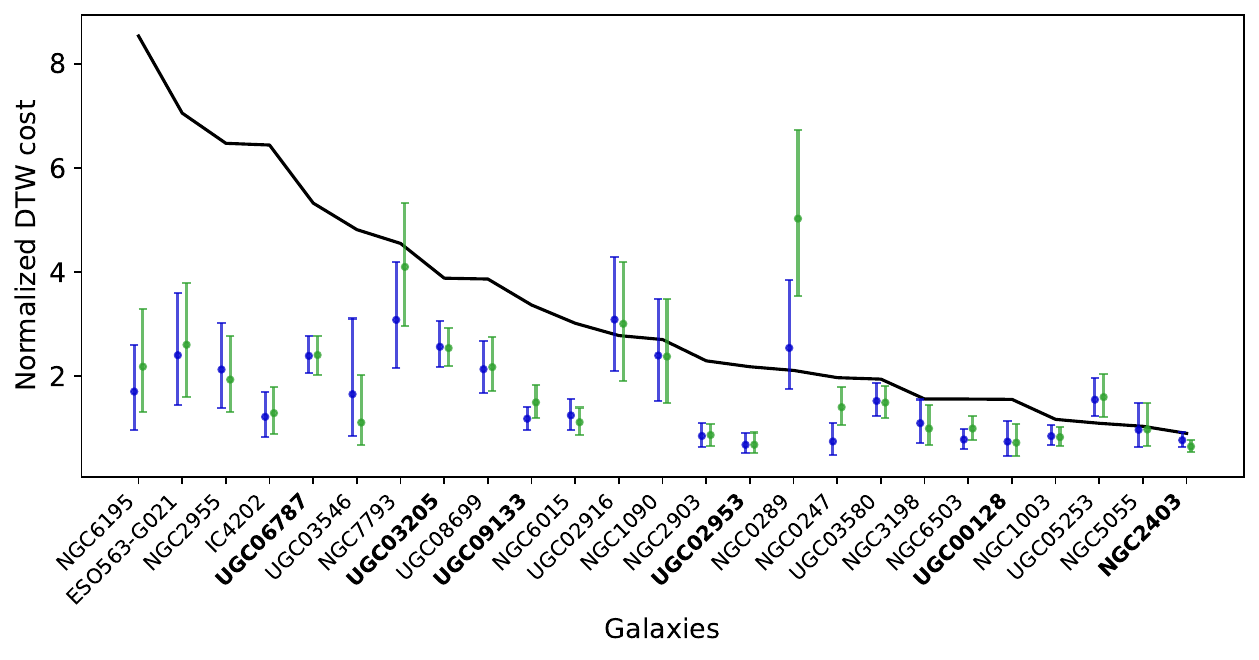}
    \end{subfigure}
        \caption{Correlation statistics of SPARC galaxies, where only data points which are part of the 31 identified features (shared among 25 galaxies) are considered. All plots are labelled identically as described in Fig.~\ref{fig:SPARC_hist}, with galaxies ordered again in ascending correlations between $\delta$\Vobs/ and $\delta$\Vbar/.
        Observe that while the expectations from MOND and \LCDM/ remain largely indistinguishable, the correlation between $\delta$\Vobs/ and $\delta$\Vbar/ is always weaker than or comparable to what is expected by both MOND and \LCDM/. Therefore, we do not find clear evidence for Renzo's rule in the SPARC dataset.}
        \label{fig:SPARC_hist_ft}
\end{figure*}

To narrow down our test for Renzo's rule, the same analysis is repeated on only the 25 galaxies where at least one feature is found in \Vobs/. Additionally, data points located outside the windows which contains such features (as identified by Algorithm~\ref{alg:ft_id}) are disregarded. Arranged similarly as before, a summary of the statistics obtained is shown in Fig.~\ref{fig:SPARC_hist_ft}.
It is clear that while the expectations from MOND and \LCDM/ remain largely indistinguishable, the correlation between $\delta$\Vobs/ and $\delta$\Vbar/ is always weaker than or comparable to what is expected by both MOND and \LCDM/.
Therefore, we find no evidence for Renzo's rule in SPARC galaxies.

Interestingly, reorganising the results presented in Fig.~\ref{fig:SPARC_hist_ft}, we find that the features identified in \Vobs/ are, on average, less correlated to \Vbar/ than is expected by both MOND and \LCDM/ with above $3\sigma$ significance\footnote{Specifically, we find an average deviation in Pearson coefficients of $-3.47\sigma$ w.r.t. MOND and $-3.03\sigma$ w.r.t. \LCDM/, and an average deviation in DTW costs of $+3.69\sigma$ (MOND) and $+3.48\sigma$ (\LCDM/).
Recall that, contrary to Pearson coefficients, positive deviations in DTW costs represent weaker-than-expected correlations, thus on average, we find features in \Vobs/ to be less correlated to \Vbar/ than expected under both sets of statistics.}.
This suggests that, contrary to Renzo's rule, we observe an excess of features (or possibly features of unexpected shapes and sizes) in \Vobs/ of SPARC galaxies, relative to that predicted by \Vbar/ under both MOND and smooth \LCDM/ haloes.
While undoubtedly surprising, there are several caveats to keep in mind when interpreting this result, which we discuss in detail under Section~\ref{sec:discussion}.

\subsection{LITTLE THINGS galaxies}
Unfortunately, only \Vobs/ is available in our dataset, and no features are identified from the 18 RCs. However, as explained below, further analysis shows that the absence of features may be the result of an inaccurate error model, which our feature identification algorithm relies on for separating features from observational noise.

Assuming the observational errors are completely uncorrelated (as is the case for SPARC), we obtained 1000 MC samples of \Vobs/ by sampling each data point from the Gaussian distribution $\mathcal{N}(V_{\text{obs}},\epsilon_{\text{obs}})$. To investigate the lack of features, we consider varying the noise threshold, $T$, with which features are identified using the algorithm in Section~\ref{sec:ft_id} (where the fiducial value was $T=2.0)$. A feature is therefore identified if a peak/trough spans at least three data points with at least one point having a height-to-noise ratio $\geq T$. We can then find the maximum value of $T$ for each galaxy at which at least one feature can be identified from the RC. Dividing by 1000, this represents the expected distribution of $T_{max}$ from pure Gaussian noise in the 18 RCs, which we can compare to our data set.

\begin{figure}
    \centering
    \includegraphics[width=\linewidth]{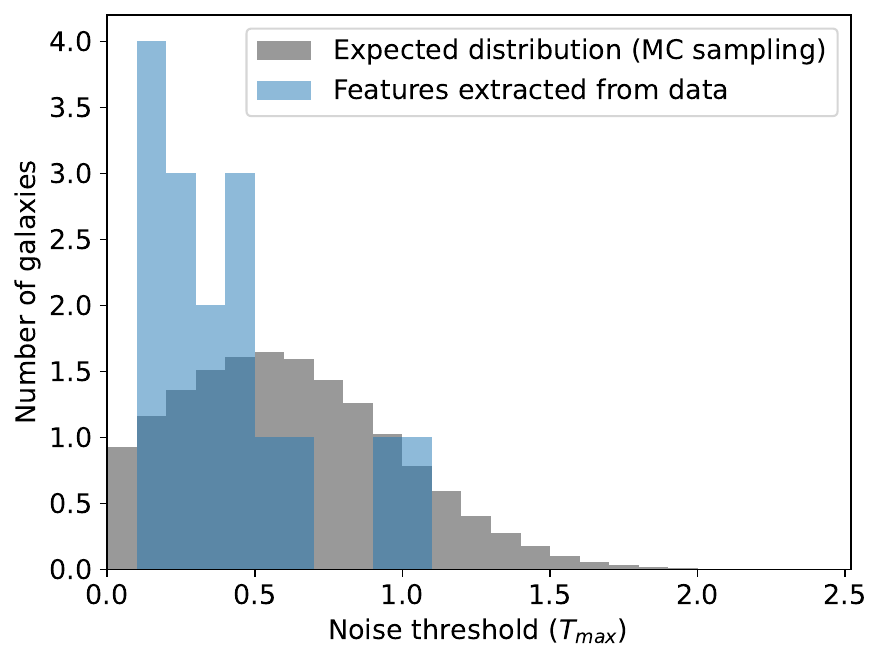}
    \caption{Histogram of the maximum noise threshold ($T_{max}$) one can impose on LITTLE THINGS galaxies such that at least one feature is identified in each corresponding RC. A comparison with MC samples shows that there are fewer features identified than what one expects from pure Gaussian noise around the same RCs, suggesting that the uncertainties in LITTLE THINGS are overestimated. This may be responsible for the apparent lack of features.}
    \label{fig:things_ft}
\end{figure}

In Fig.~\ref{fig:things_ft}, we show that our galaxies exhibit even fewer features than that expected from pure Gaussian noise, suggesting that the errors in LITTLE THINGS may be overestimated and/or correlated. 
This could explain the absence of detected features at $T=2.0$, even if such features are genuinely present.
Additionally, the expected distribution from MC sampling (which has a similar shape for SPARC) supports the choice of $T=2.0$ in Algorithm~\ref{alg:ft_id}, as fluctuations above this threshold are unlikely to arise solely from uncorrelated observational noise.

\subsection{MaGICC and CLUES simulated galaxies}\label{sec:Santos-Santos}
We performed two separate analyses for the MaGICC and CLUES simulated galaxies, one using the original simulated dataset, and one with additional scattering to estimate the effect of observational uncertainties.

\begin{figure*}
    \centering
    \begin{subfigure}{0.468\textwidth}
        \centering
        \includegraphics[width=\textwidth]{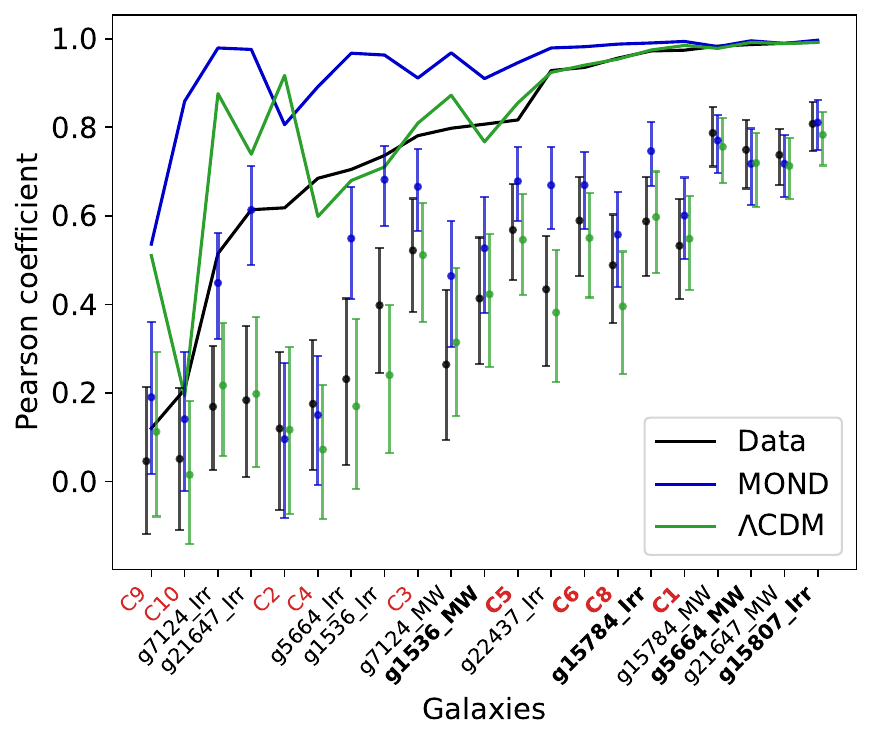}
        \caption{Pearson coefficients on MaGICC and CLUES galaxies.}
    \end{subfigure}
    \hfill
    \begin{subfigure}{0.515\textwidth}
        \centering
        \includegraphics[width=\textwidth]{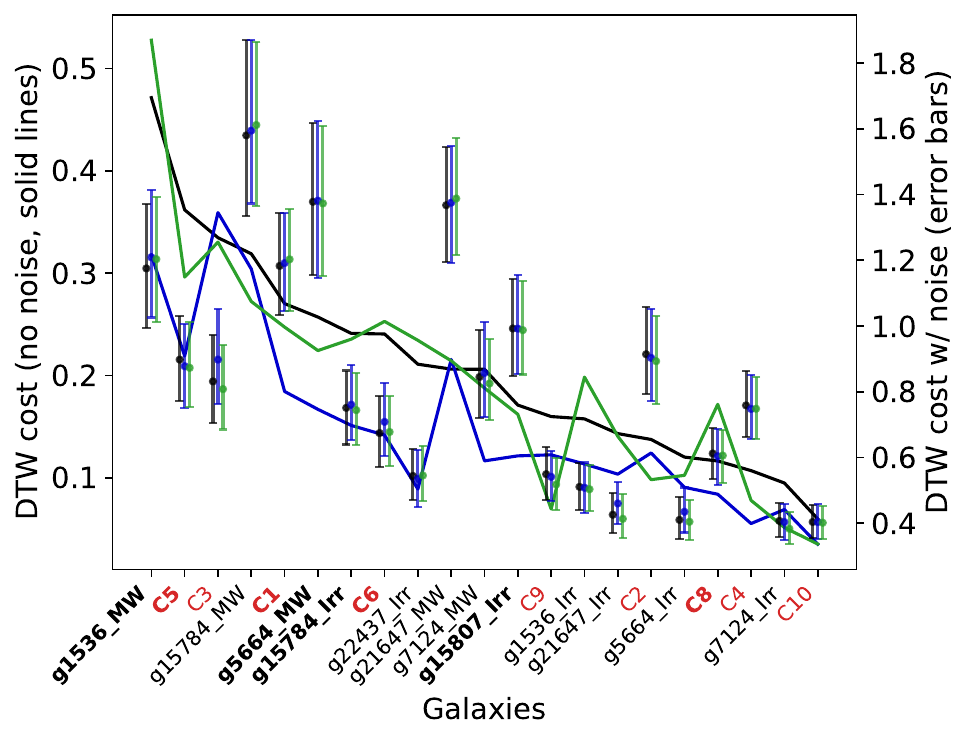}
        \caption{DTW costs on MaGICC and CLUES galaxies.}
    \end{subfigure}
        \caption{Statistics of Pearson coefficients (left) and DTW (right) on galaxies from the MaGICC (labelled along the x-axes in black) and CLUES (in red) simulations.
        In both plots, the black solid line outlines the statistics between $\delta$\Vobs/ and $\delta$\Vbar/ in ascending correlations, i.e. ascending Pearson coefficients and descending DTW costs.
        The blue and green solid lines show the statistics of $\delta$\Vmond/ and $\delta$\Vcdm/, respectively, with respect to $\delta$\Vbar/.
        Correspondingly, the error bars show the $1\sigma$ range (from MCMC fits) of correlation statistics after an artificial Gaussian noise of size 0.02$\times\max($\Vobs/) is introduced to all RCs.
        Galaxies labelled in bold along the x-axes contain noticeable features, which are (visually) identified by \protect\cite{Santos_Santos_2015} to support Renzo's rule.
        Note that to improve readability, dual y-axes are used in the DTW plot.}
        \label{fig:Santos-Santos_hist}
\end{figure*}

Using the original dataset (without observational noise), we generated \Vmond/ and \Vcdm/ using MCMC fits (see Section~\ref{sec:MCMC_fits} and \ref{sec:NGC1560_results}), and applied similar correlation statistics, the results of which are summarized as solid-line plots in Fig.~\ref{fig:Santos-Santos_hist}.
We find that, compared to expectations from MOND, both the DTW costs and Pearson coefficients suggest that features in the simulated galaxies are slightly more compatible with the weaker correlation predicted by our \LCDM/ halo fits. This indicates that the Renzo-rule behaviour reported by \cite{Santos_Santos_2015} may not be as unexpected as originally claimed, but rather exists at the level one would naively expect for a featureless halo superimposed on the baryonic mass.

To simulate systematic and observational errors in existing galaxy surveys, we introduced artificial Gaussian noise of size $0.02\times\max($\Vobs/) to all RCs; this is the average noise level present around the feature identified in Sanders's NGC 1560. The results are shown in Fig.~\ref{fig:Santos-Santos_hist}, this time with error bars outlining the $1\sigma$ ranges as obtained from MC sampling on the artificial noise.
For a handful xof galaxies, their Pearson coefficients continue to indicate closer alignment between data and \LCDM/ expectations, as opposed to that expected from MOND. However, it is clear that even with relatively conservative noise (compared to, e.g., the average scatter of $0.05\times\max($\Vobs/) in SPARC galaxies), the overlapping error bars show that, in most cases, both Pearson and DTW statistics struggle to distinguish between the MOND and \LCDM/ expectations, once again demonstrating the difficulty of testing Renzo's rule in more realistic data.

\begin{figure*}
    \centering
    \begin{subfigure}{0.468\textwidth}
        \centering
        \includegraphics[width=\textwidth]{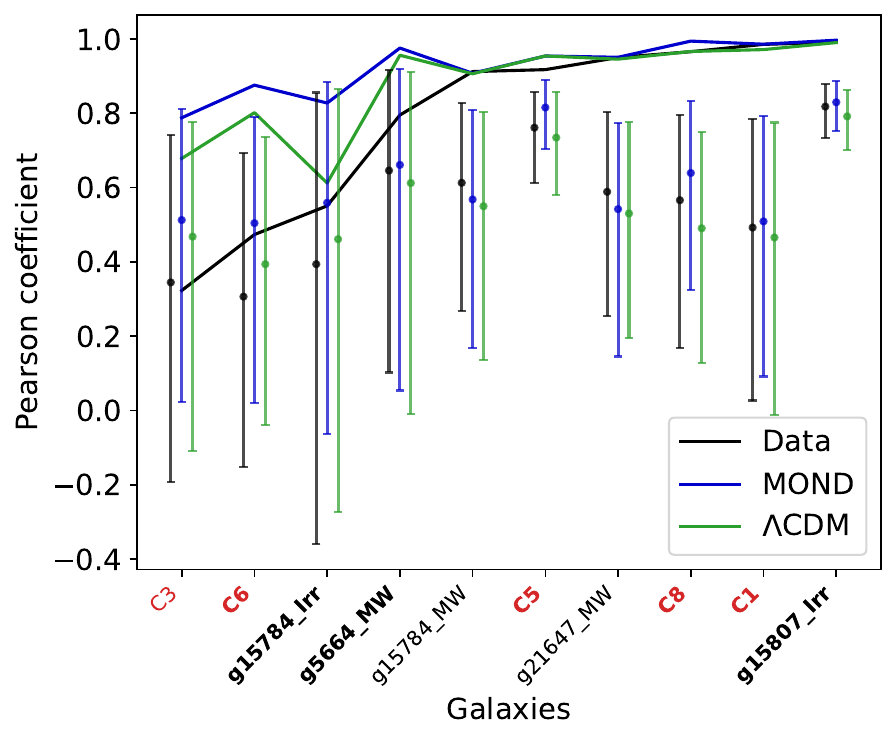}
        \caption{Pearson coefficients on MaGICC and CLUES galaxies.}
    \end{subfigure}
    \hfill
    \begin{subfigure}{0.515\textwidth}
        \centering
        \includegraphics[width=\textwidth]{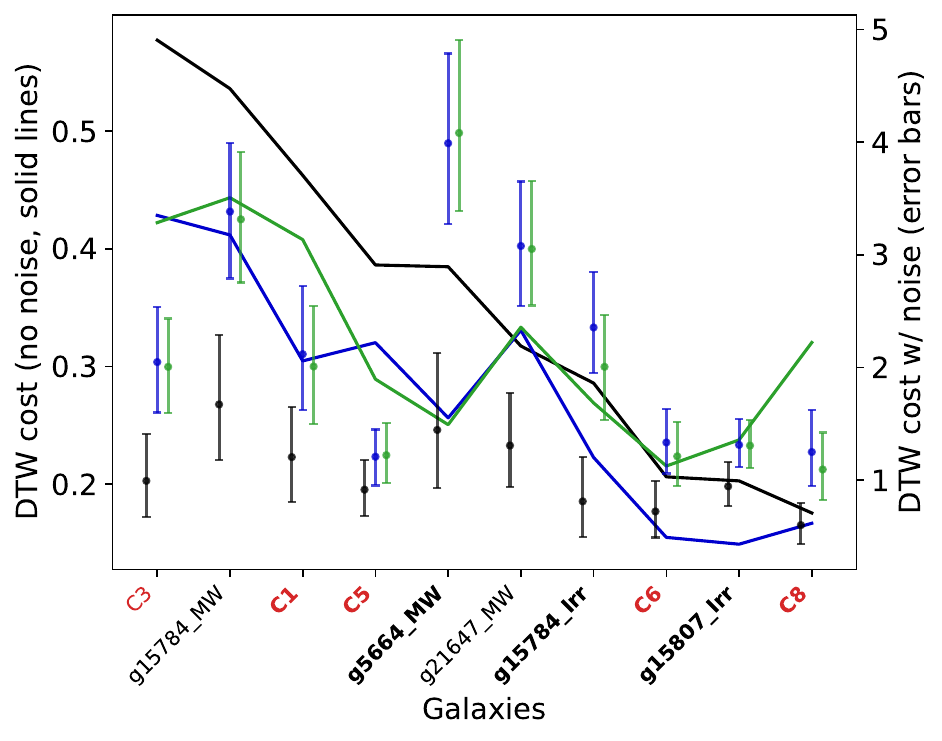}
        \caption{DTW costs on MaGICC and CLUES galaxies.}
    \end{subfigure}
        \caption{Correlation statistics for galaxies from the MaGICC and CLUES simulations, where only data points located within feature windows (as identified by Algorithm~\ref{alg:ft_id}) are considered; note that this does not necessarily coincide with the set of galaxies identified visually by \protect\cite{Santos_Santos_2015}.
        All plots are labelled and arranged in the same way as outlined in Fig.~\ref{fig:Santos-Santos_hist}.}
        \label{fig:Santos-Santos_window}
\end{figure*}

As with previous data sets, we repeat the same analyses, now restricted to data points located within feature windows identified by Algorithm~\ref{alg:ft_id}.
The results are presented in Fig.~\ref{fig:Santos-Santos_window}. Note that the set of features identified by our algorithm is slightly different to that identified visually by \cite{Santos_Santos_2015}.
Similar to before, the windowed analysis shows the data to be consistently less correlated than what is expected by MOND, and often even less than, but arguably closer to that of \LCDM/.
However, with the greatly reduced number of data points, we see that both correlation statistics are more affected by the introduction of artificial noise, with much larger error bars in both and often unstable behaviour for DTW specifically.

Given null results from both the full and windowed analyses, we conclude that no evidence for Renzo's rule is found in the MaGICC and CLUES simulated galaxies.
Specifically, we find that features in these RCs, as presented in \cite{Santos_Santos_2015}, do not correlate to a greater degree than that expected from smooth \LCDM/ haloes.

\subsection{Mock data}\label{sec:mock_data}
The preceding results indicate that currently existing data is not of sufficiently high quality to robustly identify features that correlate between the baryonic and total RCs. The culprits for this are (i) a lack of strong features in the data in the first place, and (ii) significant observational uncertainties. We are thus led to wonder what conditions would have to be satisfied for Renzo's rule to be distinctly apparent in our tests, and whether such conditions are likely to arise in future, more precise data.

\begin{figure}
    \centering
    \includegraphics[width=0.48\textwidth]{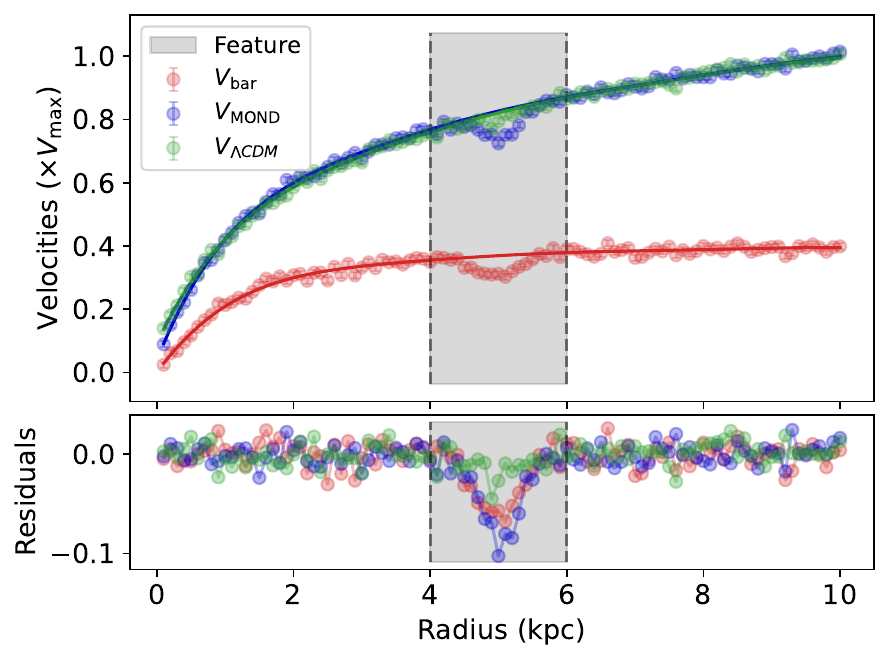}
    \caption{An example of mock data generated with sampling rate $n=10$ (such that the feature consists of $N=20$ points) and signal-to-noise ratio $h/\epsilon=4.0$. In the top panel, the error bars show the generated data (normalised by $V_{\max}\equiv\max($\Vobs/)), while the solid lines show their corresponding GPR fits, with respect to which residuals are obtained and shown in the bottom panel. The shaded region marks the 4.0-6.0~kpc window, i.e., the feature on which the windowed analysis is performed (see Section~\ref{sec:mock_data}).}
    \label{fig:mock_example}
\end{figure}

To investigate this, we perform a set of mock tests involving simplified model galaxies containing a single Gaussian feature. The procedure for this is as follows:
\begin{enumerate}
    \item Generate mock \Vbar/ using the $\arctan$ function on $r\in[0,10]$~kpc, with a variable sampling rate of $n$ data points per kpc. This is multiplied by $35\times2/\pi$ to produce a physical \Vbar/ (in km~s$^{-1}$), similar to that of NGC 1560.
    \item Imitating the feature in Sanders's NGC 1560, we subtract a Gaussian dip $-4.0\times\mathcal{N}(5.0, 0.3)$ from \Vbar/ to simulate a baryonic RC with a significant feature.
    \item Generate \Vmond/ by applying equation~\ref{eqn:MOND} to \Vbar/.
    \item Generate \Vcdm/ using MCMC fits on \Vmond/ (as outlined in Section~\ref{sec:MCMC_fits}).
    \item Scatter all RCs generated (\Vbar/, \Vmond/ and \Vcdm/) by uncorrelated Gaussian noise of variable amplitude $\epsilon$, giving 1000 sets of noisy RCs per $\epsilon$.
    \item Apply the same GPR and correlation calculations to each set of RCs as outlined in Section~\ref{sec:residuals}--\ref{sec:correlations}.
    In particular, we define $r=4.0$-6.0~kpc to be the feature window (see Fig.~\ref{fig:mock_example}), which contains $N=2n$ data points. This is such that for each RC, (a) the $N$ data points within this window are removed before applying each GPR to reduce bias towards the feature; and (b) we perform two sets of analyses for each sample: one using the full RCs, and another using only these $N$ data points.
    \item Vary the sampling rate $n$ from 2 to 40 (by integer increments) and the signal-to-noise ratio $h/\epsilon$ from 1 to 20 (by half-integer increments).
\end{enumerate}

\begin{figure*}
    \centering
    \begin{subfigure}{0.49\textwidth}
        \centering
        \includegraphics[width=\textwidth]{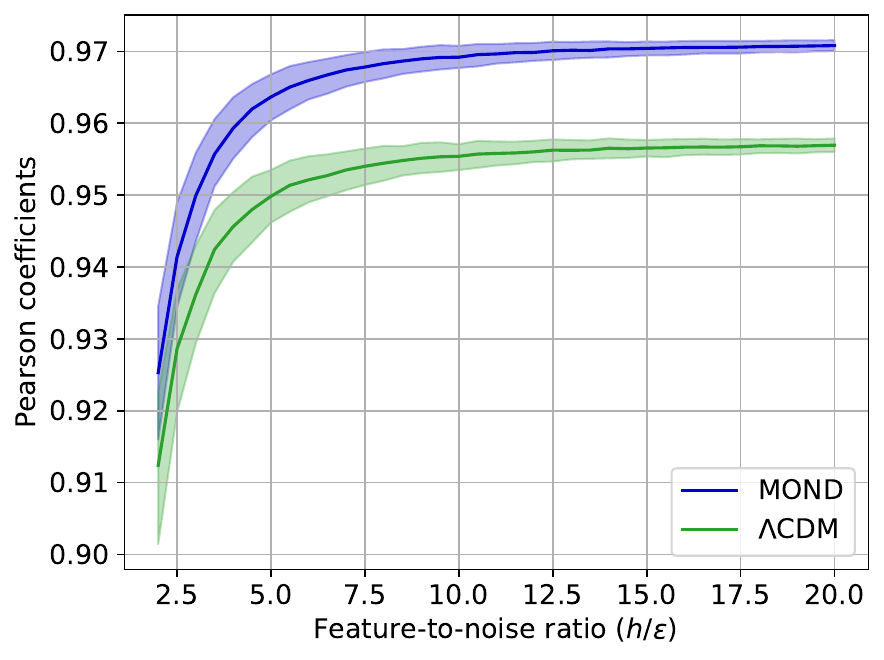}
        \caption{Pearson coefficient against feature-to-noise ratio.}
    \end{subfigure}
    \hfill
    \begin{subfigure}{0.50\textwidth}
        \centering
        \includegraphics[width=\textwidth]{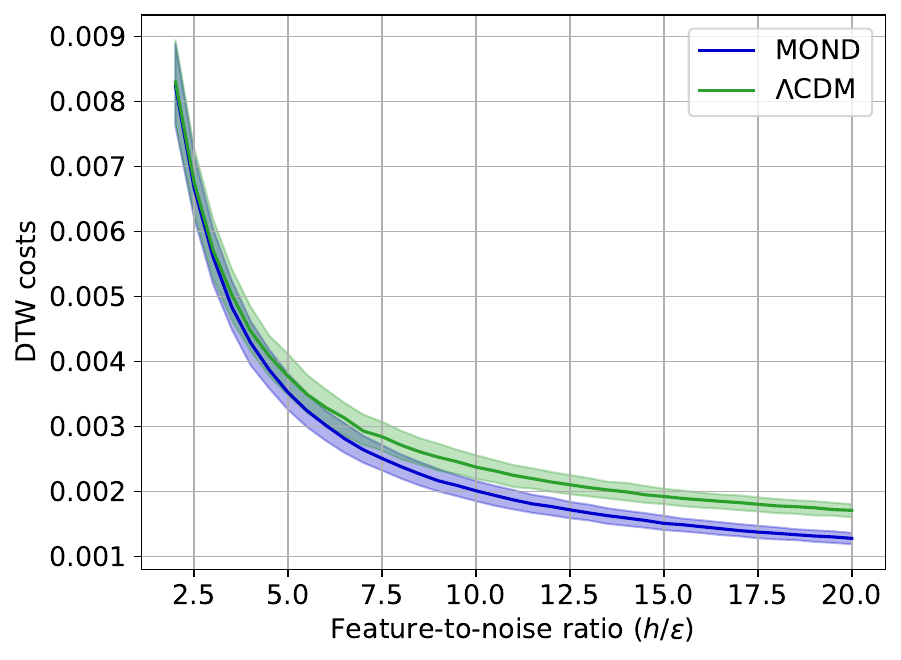}
        \caption{Normalised DTW cost against feature-to-noise ratio.}
    \end{subfigure}
        \caption{Correlation statistics at $n=10$ as a function of feature-to-noise ratio ($h/\epsilon$). Blue bands depict the (near-symmetric) $1\sigma$ spread in correlation statistics between $\delta$\Vmond/ and $\delta$\Vbar/, while the green bands show that between $\delta$\Vcdm/ and $\delta$\Vbar/.}
        \label{fig:corrVnoise}
\end{figure*}

For each sampling rate $n$, we first look at how the correlation statistics vary as a function of feature-to-noise ratio, $h/\epsilon$ (see Fig.~\ref{fig:corrVnoise}). Noting that the errors are approximately symmetric, we define the feature significance as
\begin{equation}\label{eqn:ftsig}
    S(N\equiv2n,h/\epsilon) = \frac{\vert\mu_M - \mu_\Lambda\vert}{\sqrt{\sigma_M^2 + \sigma_\Lambda^2}},
\end{equation}
where $\mu_M$ denotes the median correlation (Pearson coefficients or DTW costs) between $\delta$\Vmond/ and $\delta$\Vbar/, and $\mu_\Lambda$ denotes that between $\delta$\Vcdm/ and $\delta$\Vbar/. 
Correspondingly, $\sigma_M$ defines the near-symmetric $1\sigma$ uncertainties around $\mu_M$, while $\sigma_\Lambda$ is for that around $\mu_\Lambda$.

\begin{figure*}
    \centering
    \begin{subfigure}{0.49\textwidth}
        \centering
        \includegraphics[width=\textwidth]{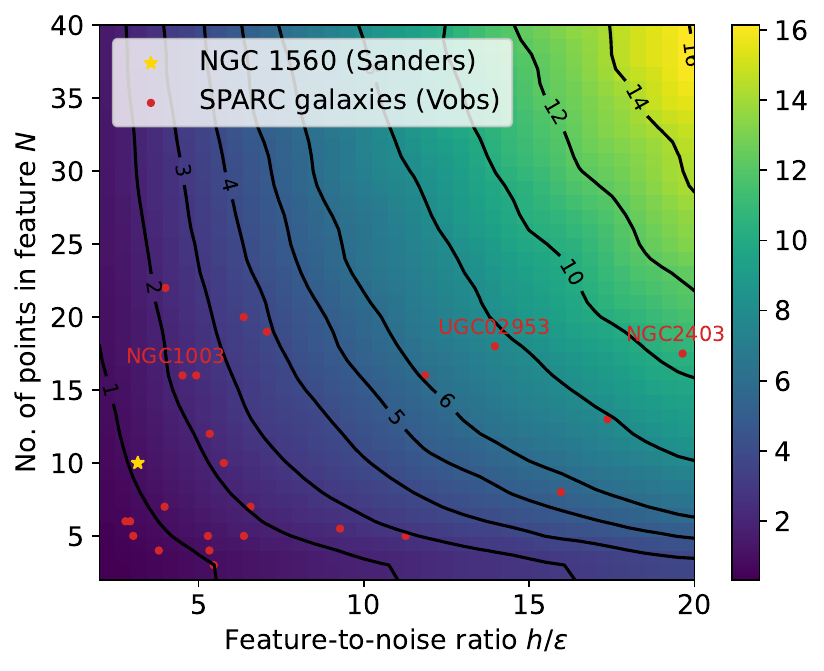}
        \caption{Significance of Pearson coefficients against $N$ and $h/\epsilon$.}
    \end{subfigure}
    \hfill
    \begin{subfigure}{0.495\textwidth}
        \centering
        \includegraphics[width=\textwidth]{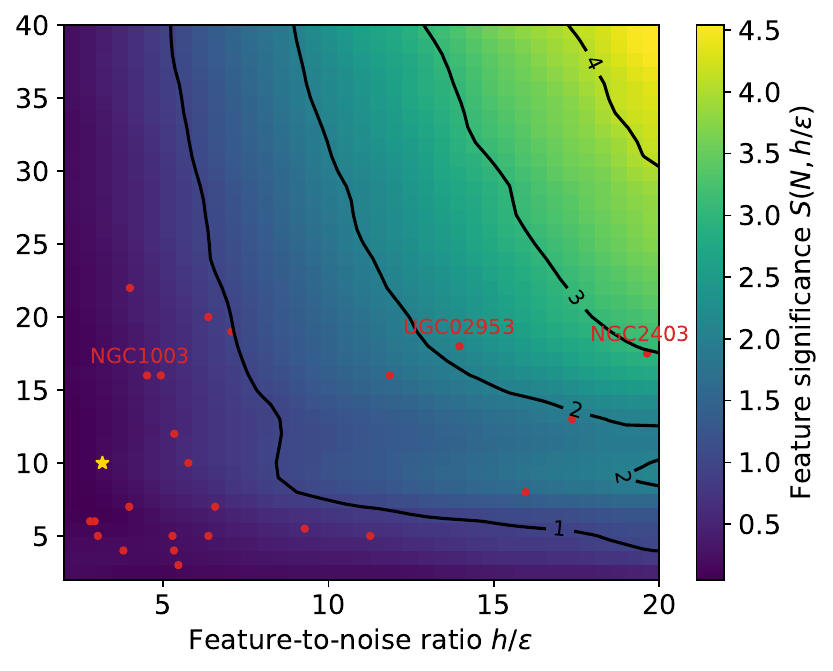}
        \caption{Significance of DTW costs against $N$ and $h/\epsilon$.}
        \label{fig:2Dsig_DTW}
    \end{subfigure}
        \caption{2D histograms of feature significance as a function of number of points in each identified feature, $N\equiv2n$, and signal-to-noise ratio, $h/\epsilon$. The significances are shown as a colour map, ranging from dark blue for $S=1\sigma$ to yellow for $\max\{S\}$. Contours are also added to highlight the boundaries of constant significance; note that only contours of even significances are shown for $S>6$ to reduce clutter.
        We also marked where some of the analysed data lie on these histograms: the feature in Sanders's NGC 1560 is marked by the gold star, and features in only \Vobs/ of SPARC galaxies are scattered as red dots.
        For SPARC, the galaxies shown in Fig.~\ref{fig:SPARC_examples} are tagged, with the exception of UGC 2953, which lies beyond the boundaries with $N=27$ and $h/\epsilon=24.2$.}
        \label{fig:2Dsig}
\end{figure*}

\begin{figure*}
    \centering
    \begin{subfigure}{0.49\textwidth}
        \centering
        \includegraphics[width=\textwidth]{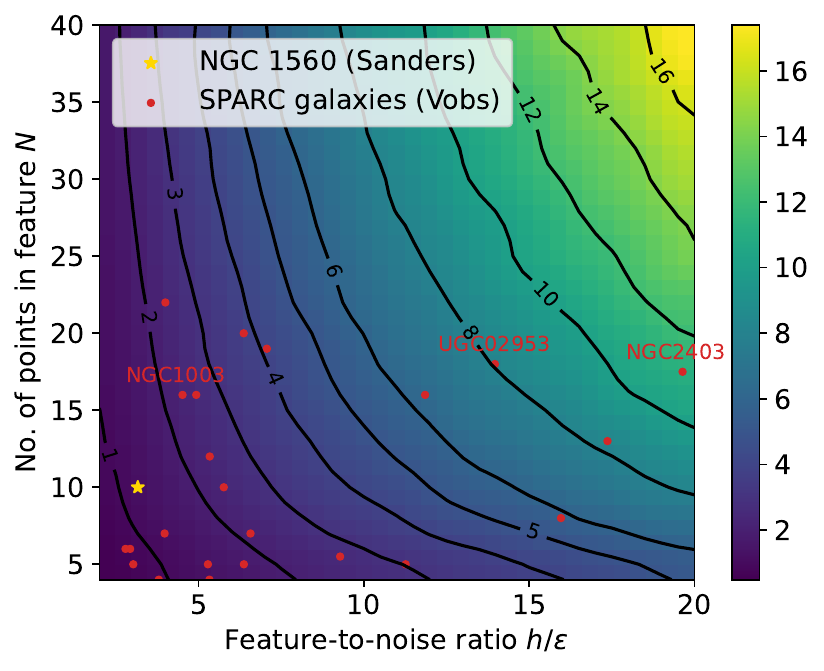}
        \caption{Significance of Pearson coefficients against $N$ and $h/\epsilon$.}
    \end{subfigure}
    \hfill
    \begin{subfigure}{0.48\textwidth}
        \centering
        \includegraphics[width=\textwidth]{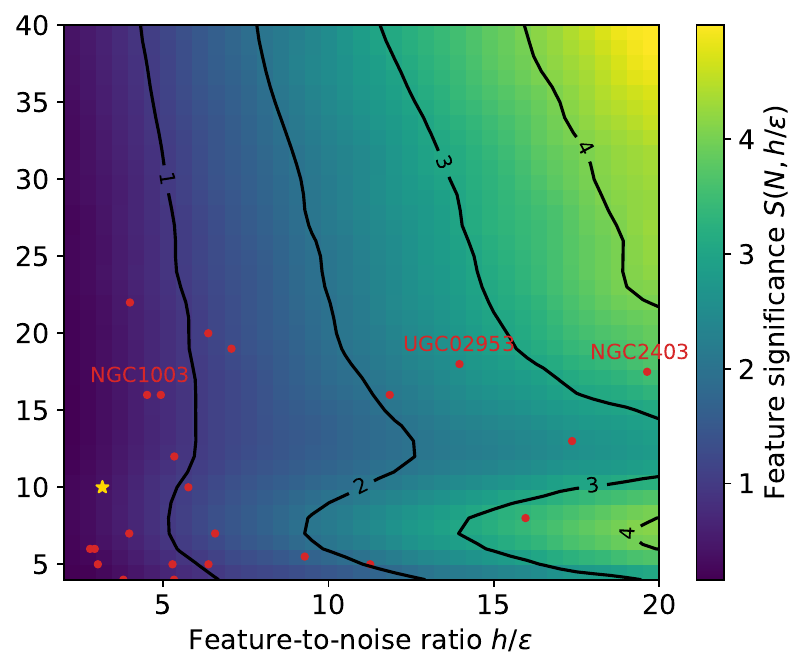}
        \caption{Significance of DTW costs against $N$ and $h/\epsilon$.}
    \end{subfigure}
        \caption{2D histograms of feature significance as a function of number of points in each identified feature, $N\equiv2n$, and signal-to-noise ratio, $h/\epsilon$. Here, significances are calculated on a window (4.0-6.0~kpc) around the artificial feature.
        All colours and labels are identical to those in Fig.~\ref{fig:2Dsig}.
        Comparing the two sets of plots, we see that focusing our analyses on the feature window does increase the significances achieved, though not as appreciably as one may naively expect.}
        \label{fig:2Dsig_window}
\end{figure*}

Using equation~\ref{eqn:ftsig}, we can further plot the feature significance as a 2-dimensional distribution over both the sampling rate $n$ and signal-to-noise ratio $h/\epsilon$.
For a Gaussian dip $\mathcal{N}(5.0, 0.3)$, the results are shown in Fig.~\ref{fig:2Dsig} (analyses with full RCs) and Fig.~\ref{fig:2Dsig_window} (analyses focused on feature window); we have verified that the overall distribution does not vary much when different feature widths (ranging from 0.2 to 0.4) are used.
As expected from earlier analyses on existing data, Pearson coefficients demonstrably outperforms DTW in distinguishing between the statistics from MOND and \LCDM/.
Comparing the two sets of histograms, we also see that the significances are generally increased when we narrow down our analyses from using full RCs (Fig.~\ref{fig:2Dsig}) to only the 4.0-6.0~kpc feature window (Fig.~\ref{fig:2Dsig_window}), though the improvement is not as appreciable as one might have expected.

Moreover, plugging in the earlier results from Sanders's NGC 1560 data to equation~\ref{eqn:ftsig}, we obtain significances of roughly $S=1.79$ for Pearson and $S=0.05$ for DTW in the full analysis, and $S=1.06$ (Pearson) and $S=0.10$ (DTW) for the windowed analysis.
Indeed, from the 2D histograms, one can expect the same feature (marked by the gold stars) to achieve significances of around $1\sigma$ for Pearson and negligible significance for DTW, in agreement with our calculations.

Furthermore, plotting the features found in SPARC, which are identified only in \Vobs/, we see that if such features exist instead in \Vbar/ with similar sampling rates and feature-to-noise ratios, they would be expected to achieve exceptionally high significances, making them incredibly useful for testing Renzo's rule; most notably, we have the `wiggle' in \Vobs/ of UGC 6787 (see bottom-right panel of Fig.~\ref{fig:SPARC_examples}), which is not shown in Fig.~\ref{fig:2Dsig} and \ref{fig:2Dsig_window} since it lies beyond the boundaries with $N=27$ and $h/\epsilon=24.2$.
This demonstrates that, if more baryonic features are present at the quality observed in \Vobs/ of SPARC galaxies, our methodology is capable of testing Renzo's rule decisively, with statistics from \LCDM/ and MOND discernible to well beyond $5\sigma$ significance.

More generally, we find that to gain higher feature significance, reducing the observational noise $\epsilon$ (such that $h/\epsilon$ increases) is often more important than enhancing the sampling rate (and therefore increasing $N$).
For instance, with the feature in Sanders's NGC 1560, which consists of 10 points and has a signal-to-noise ratio of $h/\epsilon=3.16$, one can achieve $2\sigma$ and $5\sigma$ significances with Pearson coefficients by reducing the noise by roughly $50\%$ and $80\%$, respectively; whereas, it is impossible to achieve these significances by enhancing the sampling rate alone.
Similarly, using Pearson coefficients, and with an average of $N=11.1$ for the features identified in \Vobs/ of SPARC galaxies, one can expect to test Renzo's rule with up to $2\sigma$ significance should there exist features in \Vbar/ with $h/\epsilon\approx5$, and up to $5\sigma$ with $h/\epsilon\approx12$.
Note that in present SPARC data, the average signal-to-noise ratio for features in \Vobs/ is 8.28. Hence, if a galaxy is identified to have a baryonic feature with similar properties, the method presented is a promising test for Renzo's rule up to $S=3\sigma$, with higher significances likely requiring galaxy data of higher quality.

\section{Discussion}\label{sec:discussion}
\subsection{\texorpdfstring{Renzo's Rule in \LCDM/}{Renzo's Rule in LCDM}}\label{sec:LCDM_Renzo}
Before discussing specific results, we should address the central question: is Renzo's rule compatible with the standard \LCDM/ model?
As introduced in Section~\ref{sec:intro}, Renzo's rule is often suggested as a major problem for \LCDM/ (e.g., see \citealt{Famaey_2012} and \citealt{McGaugh_2019}).
Indeed, without some strange coupling between luminous and dark matter, it seems difficult to explain the existence of strongly correlated features at supposedly DM-dominated regions of galaxies.
However, even in the absence of observational errors, there are two reasons why one may expect Renzo's rule to appear within a \LCDM/ Universe, with features resembling that expected by MOND.

Firstly, even with a perfectly smooth DM halo, e.g., one modelled by the Navarro-Frenk-White (NFW) profile (see equation~\ref{eqn:nfw}), features in \Vbar/ are still reflected in \Vobs/, often appearing larger than one intuitively expects.
Concretely, consider a DM-dominated region where \Vobs/ is around twice the magnitude of \Vbar/, such that the DM component is given by $V^2_{\text{DM}} = V^2_{\text{obs}} - V^2_{\text{bar}} = 3\times V^2_{\text{bar}}$.
Then, suppose we introduce a feature in \Vbar/ of height $h\times\text{\Vbar/}$ (e.g., $h\approx-16\%$ in NGC 1560, see Fig.~\ref{subfig:NGC1560_a}).
This generates a new \Vobs/, given by
\begin{align}\label{eqn:LCDM_expectation}
\begin{split}
    V'_{\text{obs}}
    &= \sqrt{V^2_{\text{DM}} + V^2_{\text{bar}}}
    = \sqrt{3 + (1 + h)^2} \times \text{\Vbar/} \\
    &> \sqrt{4 + 2h + h^2/4} \times \text{\Vbar/}
    = (2 + h/2) \times \text{\Vbar/}.
\end{split}
\end{align}
In particular, $V'_{\text{obs}}-V_{\text{obs}} > h/2 \times V_{\text{bar}}$, thus even in regions dominated by a perfectly smooth halo (such that $\text{\Vobs/} = 2\times \text{\Vbar/}$), one would still expect a feature in \Vbar/ to be reflected in \Vobs/ with at least half its original size.

\begin{figure}
    \centering
    \includegraphics[width=\linewidth]{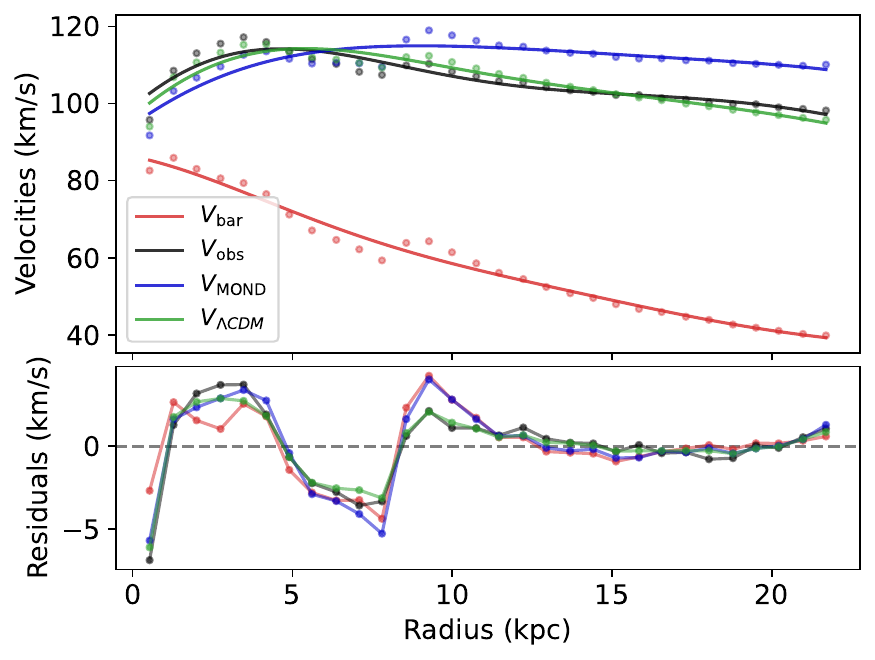}
    \caption{RCs of galaxy C5, one of the few simulated galaxies visually identified by \protect\cite{Santos_Santos_2015} to bear features resembling Renzo's rule.
    In the top panel, we show the simulated RCs (\Vobs/ and \Vbar/; black and red scatter, respectively), alongside the RCs predicted by MOND (\Vmond/; blue scatter) and \LCDM/ (\Vcdm/; green scatter) given by MCMC fits.
    A Gaussian Process Regression is applied (solid lines), which fits through the large scale, overall shape of the RCs. The residuals relative to these fits, i.e., the small-scale fluctuations on each RC, is displayed in the bottom panel.
    From this and our simple calculation with $M\approx120/60=2$, we see that the visually striking feature at around 10~kpc precisely matches the prediction from superimposing a perfectly smooth DM halo, rather than that from MOND.}
    \label{fig:C5}
\end{figure}

In fact, repeating the above calculation for a general DM halo where $V_{\text{obs}} = M \times V_{\text{bar}}$, one can show that $V'_{\text{obs}}-V_{\text{obs}} > h/M \times V_{\text{bar}}$, i.e. a feature in \Vbar/ is always reflected in \Vobs/ with at least $1/M$ times its original size.
Hence, even with a perfectly smooth DM halo, features in \Vbar/ are always reflected in \Vobs/ at a reduced, but non-zero height.
For example, we saw earlier that a handful of simulated galaxies are visually identified by \cite{Santos_Santos_2015}, which all bear strongly correlated features in \Vbar/ and \Vobs/, one of which is the galaxy C5.
In the top panel of Fig.~\ref{fig:C5}, we show the RCs of C5 as scatter plots, alongside the predictions from MOND (\Vmond/) and a smooth \LCDM/ halo (\Vcdm/; specifically one modelled by the NFW profile).
Following our general procedure, a Gaussian Process Regression is applied (solid lines), and the residuals relative to these fits, i.e., the small-scale fluctuations on each RC, is shown in the bottom panel.
We clearly see that while the residuals of \Vmond/ align well with that of \Vbar/, with almost an exact match for the bump at around 10~kpc, the same bump in the residuals of \Vobs/ actually replicates that of \Vcdm/, both at approximately half the height of their counterpart in \Vbar/.
Indeed, by plugging in $M\approx120/60=2$ into our calculation above (and following the exact computation shown in equation~\ref{eqn:LCDM_expectation}), we see that, perhaps counter-intuitively, the visually striking feature is nothing but precisely expected from the introduction of a perfectly smooth DM halo.
This key argument applies to many of the features we identify, especially where features are located in less DM-dominated regions with $M<2$.

Secondly, baryons do interact with DM gravitationally, leading to substructures in DM haloes which emerge as correlated features in RCs.
For instance, a recent study by \cite{Bernet2025} shows that in a range of state-of-the-art \LCDM/ simulations, the introduction of stellar spiral arms generates clear spiral-shaped overdensities in its otherwise smooth DM halo. This gravitational back-reaction may reflect as strongly correlated features in \Vbar/ and \Vobs/, even in DM-dominated regions of RCs, recreating Renzo's rule in a purely \LCDM/ Universe.
Indeed, it would be interesting to apply our analysis on similar numerical simulations in the future.

\subsection{Interpretation of results}
For the first time, we provide a systematic and statistical test of \emph{Renzo's rule} for RCs in both observed and simulated galaxy data.
In particular, by considering residuals from Gaussian Process Regressions (GPR), we compute the Pearson coefficients and dynamic time warping (DTW) costs of small-scale features in \Vobs/ and \Vbar/, which are compared to that predicted by Monte-Carlo samples of MOND and \LCDM/; specifically, we assume the simple interpolation function for MOND (see equation~\ref{eqn:simpleIF}), and smooth Navarro-Frenk-White (NFW) haloes (see equation~\ref{eqn:nfw}) for \LCDM/.

For the dwarf spiral NGC 1560, we find hints of Renzo's rule in datasets from both \cite{Sanders_2007} and \cite{Gentile_2010}, with a clear feature identified in RCs of the former.
Notably, the statistics obtained from both datasets deviate from MOND and \LCDM/ expectations, with up to $5\sigma$ significance for the RCs from Gentile et al., implying an exceedingly strong correlation of features which seems incompatible with both models.
However, we note that points within the 4.2-6.2~kpc window of RCs from Gentile et al. are removed for the GPR, despite our algorithm finding no features with the original fit.
While well-motivated by the features identified in Sanders's dataset, this likely introduced bias into the statistics obtained by amplifying the residuals within said window.
Numerically, note that our analysis only separates the predictions from MOND and \LCDM/ to a statistical significance of $S=1.79\sigma$ at best, with below $1\sigma$ in all analyses on RCs from Gentile et al., thus the results obtained are more likely a reflection of overestimated and correlated observational errors in the data, which our feature identification algorithm (Algorithm~\ref{alg:ft_id}) and Monte-Carlo methods are sensitive to, rather than a clear preference or rejection of either model.

Moreover, the stark difference in both shapes and sizes of the features found between the two datasets is concerning.
Upon further investigation, we find that Sanders's RC resembles only the northern (receding) side of the RCs modelled by Gentile et al. (see fig.~9 and 13 in \citealt{Gentile_2010}).
While this does not invalidate Sanders's data as an interesting test for Renzo's rule, the reality of the clear feature identified is disputable.
This calls for more refined data of the dwarf spiral, specifically surveys with better-controlled observational uncertainties, before any conclusive claims can be established.
Additionally, as will be discussed in Section~\ref{sec:future_work}, improvements should be made to incorporate strongly correlated uncertainties in the data.
Furthermore, these (almost always) non-axisymmetric features also suggests that applying Renzo’s rule to a galaxy’s full two-dimensional velocity field -- along with its corresponding two-dimensional baryonic distribution -- may offer deeper insights than analyses based solely on azimuthally averaged RCs.

In other galaxy surveys, we find 25 galaxies from the SPARC dataset containing distinct features in their total RC (\Vobs/), but no features are found in any of their baryonic counterparts (\Vbar/).
Similarly, we also find no features in any of the galaxies from the LITTLE THINGS survey.
For the latter, a simple analysis suggests that features are likely obscured by overestimated errors, which our feature identification algorithm (Algorithm~\ref{alg:ft_id}) relies on.
On the other hand, while for most SPARC galaxies, the lack of baryonic features is due to a similar overestimation of (strongly correlated) errors, there are features in \Vobs/ which bear no resemblance in \Vbar/ (see Fig.~\ref{fig:SPARC_examples}).
These features lead to correlation statistics which are lower than what is expected by both MOND and smooth \LCDM/ haloes, and with an average deviation of above $3\sigma$, they appear incompatible with both models.
In particular, this result contradicts Renzo's rule, which states that for each feature in \Vobs/, there should be a feature in \Vbar/ (and vice versa).

In fact, upon closer inspection, this problem of excess features in \Vobs/ of SPARC galaxies has long been present in the literature, albeit not explicitly mentioned.
For example, studies which fit these RCs with the radial acceleration relation (RAR; \citealt{Li_2018}), alternative DM haloes (e.g., Einasto halo fits by \citealt{Amir_2019}), even self-interacting DM \citep{Tao_2019} and modified gravity theories \citep{Das_2023}, all display fits which struggle to reproduce small-scale features in \Vobs/, often involving some of the galaxies presented in Fig.~\ref{fig:SPARC_examples}.
Indeed, a solution to this problem will likely require further in-depth analyses of individual problematic galaxies.
That said, our analysis should serve as an interesting test for both new and existing astrophysical theories, which ought to allow sufficient flexibility (e.g., via local variations in DM densities) to generate similar features in \Vobs/ (without necessarily correlated features in \Vbar/), while reproducing certain ``summary" statistics of tight baryon-DM relations such as the limited intrinsic scatter of the RAR and the baryonic Tully-Fisher relation.

As for simulated data, we turn to galaxies from the MaGICC and CLUES hydrodynamical \LCDM/ simulations. These datasets are of particular interest since \cite{Santos_Santos_2015} visually identified clear features in the RCs of 9 galaxies, resembling Renzo's rule without its explicit implementation, nor that of any closely related phenomena such as the baryonic Tully-Fisher relation.
While this appears surprising, we find no evidence that features in these galaxies correlate at a stronger level than is expected by smooth \LCDM/ haloes; in fact, the statistics obtained often agree more with that predicted by \LCDM/ than of MOND.
Moreover, contrary to intuitive expectations, a simple calculation shows that even with the addition of a perfectly smooth halo (such as one modelled by the NFW profile), features in \Vbar/ are expected to reflect in \Vobs/ at $1/M \equiv V_\text{bar}/V_\text{obs}$ times its original height, which appears to be the case for some of the identified features (see Section~\ref{sec:LCDM_Renzo}).

Using RCs modelled by $y=\arctan x$ with an additional Gaussian dip, our analyses with mock data (Section~\ref{sec:mock_data}) gives a wider perspective of the statistics obtained, and what improvements are needed in future galaxy data for a decisive test of Renzo's rule, specifically one which distinguishes the expected statistics from MOND and smooth \LCDM/ haloes to higher significance.
Fig.~\ref{fig:2Dsig}-\ref{fig:2Dsig_window} show that for Pearson coefficients, it is often more important to reduce noise (thus increasing the feature-to-noise ratio) than enhancing sampling rates for data lacking in both aspects (as is the case for NGC 1560).
However, we see that there are features found in \Vobs/ of SPARC galaxies, such as UGC 2953, NGC 2403 and most notably, UGC 6787, which, if they exist in \Vbar/ instead, would distinguish expectations from MOND and \LCDM/ to far beyond $5\sigma$ significance.
Unfortunately, the baryonic feature in NGC 1560 remains the most promising candidate (and indeed one of the most seen in presentations of Renzo's rule), and with our current methods, even with a 4-fold increase in sampling rate, at least a $60\%$ reduction in noise is required to achieve a potentially conclusive $5\sigma$ significance (or around $80\%$ if the sampling rate is not drastically enhanced).

A clear issue we have yet to discuss is the lacklustre performance of DTW, as shown in Fig.~\ref{fig:2Dsig}-\ref{fig:2Dsig_window}. As opposed to Pearson coefficients, contours for $S=5\sigma$ do not even appear on the histograms for DTW.
While disappointing, we emphasize that DTW is a flexible overarching framework, and our algorithm defined in Section~\ref{sec:dtw} is only one of the simplest versions available.
For instance, one can define different DTW metrics to that of equation~\ref{eqn:dtw}, apply larger weights on identified features, modify DTW to account for local gradients and shapes, etc. (for a summary and comparison of different DTW variants, see, e.g., \citealt{Abdelmadjid_2021}).
Therefore, our study does not completely rule out DTW as a tool for testing Renzo's rule, though modifications of which are evidently required for obtaining significant results.

\subsection{Caveats and potential improvements}\label{sec:future_work}
We should once again emphasize that in all results presented, MOND refers specifically to equation~\ref{eqn:MOND} together with the simple interpolating function (equation~\ref{eqn:simpleIF}), while \LCDM/ refers only to smooth DM haloes modelled by the NFW profile (equation~\ref{eqn:nfw}).
In particular, for MOND, \cite{Milgrom_2008} showed that certain families of interpolating functions (IFs) can generate shell-like features at specific transition radii, even in the absence of any corresponding structure in \Vbar/. These features manifest as “phantom dark matter”, resembling the isolated features identified in \Vobs/ of some SPARC galaxies.
While such presentations often prefer IFs with a very fast MOND-Newtonian transition -- contrary to the gradual transition preferred by the SPARC dataset -- it is clear that our study does not invalidate MOND in general. Rather, this challenges the validity of the Simple or RAR IF and may even indicate the need for the modified inertia interpretation of MOND; all of these are discussed in detail by \cite{Desmond:2024eic, moriond}.
Future tests of Renzo's rule should therefore allow for variation of the IF.

For \LCDM/, by not incorporating any local gravitational interactions between baryons and DM (e.g., spiral arms in \citealt{Bernet2025}), we expect our results to underestimate the correlation statistics expected by more sophisticated \LCDM/ calculations, although it remains unlikely that such improved statistics would exceed the existing predictions from MOND.
Furthermore, by assuming all DM haloes to be smooth and featureless, it is difficult to explain features in \Vobs/ which are not reflected in \Vbar/, such as those found in the SPARC dataset.
Hence, both our $\Lambda$CDM and MOND implementations are inherently smooth, and exclude the small-scale physics required to generate features in \Vobs/ in addition to those reflected directly from \Vbar/.
Of course, limitations of the (idealised) error models of the data, e.g., the disregard of asymmetries, may also lead to unmodelled features.

Throughout our analysis, we have also assumed that all stellar orbits are circular in galaxies. This is a simplification of the true physics involved -- it is likely that non-circular motion, e.g., in barred galaxies, can generate RC fluctuations that do not accurately represent true features in the underlying gravitational field (see \citealt{Milgrom_2008} for a more detailed discussion).
That said, by comparing to \LCDM/ hydrodynamical simulations, \cite{Santos_Santos_2015} points out that most RCs should not be significantly altered by assuming a spherical potential and considering the average mass of spherical shells.
Indeed, in most stable, late-type galaxies, such as those contained in the SPARC database, non-circular motions only manifest as velocity dispersions, rather than significant asymmetries across the disc.
Thus, our analysis remains a useful first step towards understanding the physics of small-scale galaxy features.

Based on the methods presented, improvements can still be made in the future without the use of sophisticated modelling or simulations.
A common problem in present data is the strongly correlated errors from uncertainties in galaxy parameters, such as distances, inclinations and mass-to-light ratios.
This makes it difficult to identify features in RCs, especially in \Vbar/, as is clear from our analyses on SPARC galaxies.
To tackle this, it may be necessary to develop new algorithms to extract RCs with more precise uncertainty quantification, including \HI{} surface densities, velocities and correlations between the various galaxy parameters.
This new algorithm should also contain flexible enough models to accurately extract features in both the surface densities and RCs.
Moreover, part of the uncertainties in present data originate from the averaging of approaching and receding sides of each galaxy -- indeed, as discussed previously, this is likely responsible for the difference between the RCs extracted by \cite{Sanders_2007} and \cite{Gentile_2010}.
It would be beneficial for future compilations to provide the two sides separately as two sets of RCs, such that our methodology can be applied independently to the better-controlled, and possibly more abundant features.

Alternatively, the problem can also be tackled by modifying Algorithm~\ref{alg:ft_id}, where instead of treating each data point separately, we evaluate segments of the normalized residual using alternative metrics such as the Mahalanobis distance \citep{Mahalanobis}; this takes into account the covariance of uncertainties, which in turn can be estimated using existing Monte-Carlo realizations from the generation of \Vmond/ and \Vcdm/.
Note that the covariance structure can vary significantly between different galaxies, and even between different RC components within the same galaxy -- indeed, when using Mahalanobis distances, this variability makes it difficult to define a consistent threshold (analogous to $T$ in Algorithm~\ref{alg:ft_id}) for systematic feature identification, even when allowing separate thresholds for \Vbar/ and \Vobs/.
To address this, one may need to modify the covariance matrices, e.g., by adding small diagonal terms or assuming a minimum uncertainty at each data point, such that a uniform threshold can be applied across more comparable covariances.
Of course, to further improve the reliability and accuracy of the algorithm (and related components like the GPR), one could always employ advanced machine learning techniques, such as those frequently used in high energy physics (e.g., for a potentially suitable Python-based algorithm, see \citealt{pyBumpHunter}).

Similarly, there are plenty of directions to pursue for improving the obtained correlation statistics.
Apart from using other variants of DTW, as previously mentioned, one could, for example, investigate the 1st and/or 2nd derivatives of the RC residuals.
In fact, as part of this study, we did examine the derivatives of RCs using spline fits --- a particularly simple but promising technique we came across is the Piecewise Cubic Hermite Interpolating Polynomial (PCHIP).
However, due to various challenges such as difficulty in physically interpreting the statistics obtained, we did not pursue this direction fully.
That said, future studies can certainly utilize similar techniques to extract more information from both existing and future galaxy data; indeed, it is feasible that one could achieve more significant results by clever manipulation of these higher derivatives alone, without drastic changes to the formulation of Pearson coefficients and DTW presented.

Of course, with all being said, the most obvious path to achieving a definitive test of Renzo's rule is to bring in galaxy data with much larger baryonic features than those in present data (e.g., comparable to those found in \Vobs/ of SPARC galaxies), preferably with small and well-modelled uncertainties, along with high sampling rates around such features.

\section{Summary}\label{sec:summary}
In this paper, we present a systematic statistical analysis for \emph{Renzo's rule}, an informal astrophysical phenomenon which until now relied solely on visual inspections of rotation curves.
Using residuals from Gaussian Process Regression (GPR), a simple feature identification algorithm and correlation quantification with Pearson coefficients and dynamic time warping (DTW), we analyse four sets of galaxy data: the dwarf spiral NGC 1560, the SPARC database, the LITTLE THINGS survey, and simulated data from the MaGICC and CLUES \LCDM/ simulations.

We identify features in NGC 1560 -- a prime example for Renzo's rule and MOND -- where correlation statistics support Renzo's rule with a slight preference for MOND over smooth \LCDM/ haloes.
However, all other results suggest that Renzo's rule is not a clear facet of galaxy phenomenology; notably, features identified in RCs of the SPARC dataset correlate around $3\sigma$ less on average with their baryonic counterparts than is expected by both MOND and \LCDM/, challenging the validity of Renzo's rule.
Thus overall, we do not find clear evidence that correlations are present to a greater degree than would be expected in a $\Lambda$CDM galaxy formation scenario.

To estimate the data quality required for a robust test of Renzo's rule, we also generate and analyse mock data with features of variable sampling rates and signal-to-noise ratios.
We find that in general, Pearson coefficients outperform DTW in differentiating MOND and \LCDM/ statistics;
and to achieve significant separation between MOND and \LCDM/ expectations, it is often more important to reduce observational uncertainties than to enhance the sampling rate.
Furthermore, by overlaying features identified in the SPARC dataset, we demonstrate that our methodology is capable of testing Renzo's rule decisively, with MOND and \LCDM/ statistics distinguishable beyond $5\sigma$ significance.
Therefore, the lack of features in baryonic rotation curves in present data -- partly due to large and correlated uncertainties in galaxy parameters -- remains the single most important barrier from a conclusive test of Renzo's rule.

\section*{Data availability}
Data from the SPARC galaxy survey is publicly available at \href{http://astroweb.cwru.edu/SPARC/}{\texttt{http://astroweb.cwru.edu/SPARC/}}. All other data and computer code underlying this study will be made available upon reasonable request to the corresponding author.

\section*{Acknowledgements}
We thank Fedir Boreiko, Benoit Famaey, Moti Milgrom, Anastasia Ponomareva, Stacy McGaugh and Nathaniel Starkman for useful inputs and discussion.

EK, TY and MJ acknowledge support from UKRI Frontiers Research Grant [EP/X026639/1], which was selected by the ERC.
EK also acknowledges support from the Oxford University Astrophysics Summer Research Programme.
HD is supported by a Royal Society University Research Fellowship (grant no. 211046).
RS acknowledges financial support from STFC Grant No. ST/X508664/1 and the Snell Exhibition of Balliol College, Oxford.

We thank Jonathan Patterson for smoothly running the Glamdring Cluster hosted by the University of Oxford, where the data processing was performed.

\bibliographystyle{mnras}
\bibliography{main}

\bsp
\label{lastpage}
\end{document}